\documentclass[preprint,showpacs,preprintnumbers,amsmath,amssymb]{revtex4}

\usepackage{graphicx}
\usepackage{bm}

\newcommand\Tm{\langle\mathbf{T}\rangle} 
\newcommand\Tmm[1]{\langle\mathbf{T}_{#1}\rangle} 
\newcommand\Tmr[1]{\langle\mathbf{T}({\bf r}_{#1})\rangle} 
\newcommand\Tmur[2]{\langle\mathbf{T}^{#1}({\bf r}_{#2})\rangle}
\newcommand\Tmmr[2]{\langle\mathbf{T}_{#1}({\bf r}_{#2})\rangle}
\newcommand\K[3]{q_{#1}(#3)(H(#2|{\bf i}_{#1}(#3))-
H(#2|{\bf r}_{T_{#1}}))}
\newcommand\Kx[2]{q_{#1}(#2)(H^*({\bf i}_{#1}(#2)|{\bf r}_{T_{#1}})-
H^*({\bf r}_{T_{#1}}|{\bf r}_{T_{#1}}))}
\newcommand\lap[1]{\left< e^{-s\mathbf{T}} \right>_{#1}}

\begin{document}

\title{Random walks and Brownian motion: a method of computation for 
first-passage times and related quantities in confined geometries}

\author{S. Condamin}
\author{O. B\'enichou}
\author{M. Moreau}
\affiliation{Laboratoire de Physique Th\'eorique de la Mati\`ere Condens\'ee 
(UMR 7600), case courrier 121, Universit\'e Paris 6, 4 Place Jussieu, 75255
Paris Cedex  }

\date{\today}

\begin{abstract}
In this paper we present a computation of the mean 
first-passage times both for a random walk in a discrete bounded lattice, 
between a starting site and a target site, and for a Brownian motion in a 
bounded domain, where the target is a sphere. 
In both cases, we also discuss the case of two 
targets, including splitting probabilities, and conditional mean first-passage
times. In addition, we study the higher-order moments and the full 
distribution of the 
first-passage time. These
results significantly extend our earlier contribution 
[Phys. Rev. Lett. {\bf 95}, 260601]. 
\end{abstract}


\maketitle

\section{Introduction}

The time it takes to a random walker to go from  a starting site to a
target site, the so called first-passage time (FPT), is an especially
important quantity that underlies a wide range of physical
processes\cite{VanKampen,Redner}.  Indeed, numerous real situations,
such as diffusion limited reactions  \cite{Rice} or animals searching
for food \cite{nous},  can be rephrased as first-passage problems. In
all these situations, the FPT is a limiting factor. As a consequence,
it is crucial to determine how this quantity depends on the parameters
of the problem.

Among these parameters, geometrical factors turn out to be
determining. For example, the mean first-passage time (MFPT) between a
starting site and a target site for a 2D random walker is infinite if
the walk is not bounded. On the contrary, it becomes finite as soon as
the walk is confined. But how does the MFPT depend on the confining
surface? In fact, the answer to this general question appears as a
difficult task, because explicit determinations of FPT are most of the
time limited to  very artificial geometries, such as 1D and
spherically symmetric problems \cite{Redner}.
 
However,  in most of the real situations, the searcher performs a
random walk in more general confining  geometries. This is for example
the case  in biology, where biomolecules often follow  a complicated
series of transformations, which are located at precise parts  of the
cell. Determining the influence of the shape of the cell on  the FPT
actually appears as a first step in the understanding of the  global
kinetics of the process.

This question of determining first passage properties in general confined geometries has raised a growing attention (see for example 
\cite{henry1,henry2,henry3,ferraro1,ferraro2,blanco,mazzolo,benichou05,condamin,LevitzJPCM05,LevitzPRL06}). Two important results  have notably
been obtained. First, in  the case of discrete random walks, an
expression for the mean first passage  time (MFPT) between two nodes
of a general network has been found \cite{Noh}. However, no quantitative 
estimation of the MFPT was derived in this paper.  
Second, the leading behavior of  MFPT  of a continuous
Brownian motion at a small absorbing window of a general reflecting
bounded domain  has been given \cite{Holcman05,bere}. These studies have 
even been extended
to a situation with a deep  potential well, leading to a generalization
of the Kramers formula \cite{Langer}.  In the case
when this window is  a small sphere within the domain, the  behavior
of MFPT has  also been derived \cite{Pinsky}. This result is
rigorous, but does not give access to the dependence of the MFPT with
the starting site.

Very recently \cite{CondaminPRL}, we have proposed a novel 
approach which allowed us to
propose accurate estimations of first passage times of {\it discrete}
random walks in confined geometry. Preliminary  results  concerning a
continuous Brownian motion have also been announced.  The main purpose
of this paper is to provide   a detailed analysis of this {\it
continuous} case, relevant to many real physical situations. In
addition, we  extend  our previous work in several directions, for
both discrete and continuous cases: the complete distribution of
FPTs is obtained; extra quantities, as conditional MFPT
in the case of several targets or mean exit times
by  a small aperture of a general reflecting bounded domain, are
derived.

The paper is structured as follows. In Sec. \ref{discret}, we first present the
computation method of FPTs in the case of  random walks
on discrete lattices.  This study includes the obtention of the MFPT,
 a comprehensive derivation of the expression of
the higher order moments as well as the complete distribution of the
FPT, whose physical meaning is extensively analysed. 
The situation with two competitive targets is also
studied, and we compute MFPT, splitting probabilities, and conditional 
MFPT. 

 In Sec. \ref{continu}, we extend all these results to the case of a
continuous Brownian motion, and detail the specific difficulties encountered
in this case. 

The explicit results obtained in Secs. 
\ref{discret} and \ref{continu} 
involve pseudo-Green functions of  a Laplace type operator,
with given boundary conditions. The Appendix \ref{approximations} is 
devoted to the evaluation of these pseudo-Green functions. For several domain
shapes, an exact formula can be obtained, which gives, for the 
quantities computed in the article, exact explicit expressions in the 
discrete case, or accurate approximations in the continuous case. 
For other domain shapes, basic approximations are proposed. 

These results are briefly summarized in Sec. \ref{discussion}, with 
a discussion of the important parameters to take into account and of the 
qualitative behavior of the MFPT

\section{Random walks on discrete lattices\label{discret}}

\subsection{Mean first-passage time}

Let us consider a point performing a random walk on an arbitrary bounded 
lattice with reflecting boundaries. We want to compute the MFPT $\Tm$ of the 
random walker at target site $T$, starting from a 
site $S$ at time $0$. We summarized this computation
in a previous article 
\cite{CondaminPRL}.

However, since it is the basis of all the developments 
explained in this article, we found useful to give it here in full detail, 
with the addition of several necessary precisions. 

Our method is based on a formula given by Kac \cite{Aldous},
concerning irreducible graphs, such that any point can be reached from 
any other point. An irreducible graph admits a unique
stationary probability $\pi({\bf r})$ to be at site ${\bf r}$ (physically, 
this is the probability for a particle which has been in the domain for a 
long time to be at site ${\bf r}$. If the transition probabilities are 
symmetric this stationary probability is uniform.). We
consider random walks starting from an arbitrary point of a subset
$\Sigma$ of the lattice, chosen with probability $\pi({\bf r})/\pi(\Sigma)$,
where $\pi(\Sigma) = \sum_{{\bf r} \in \Sigma} \pi({\bf r})$. Then, Kac's 
formula asserts that the mean number of steps needed to return to any point
of $\Sigma$, i.e. the mean first-return time (MFRT) to $\Sigma$ is 
$1/\pi(\Sigma)$. A simple proof 
of this result and of its extension to higher-order moments, which will be used
later on, is given in Appendix \ref{AnnKac}.

\begin{figure}[h!]%
\centering\includegraphics[width = .7\linewidth]{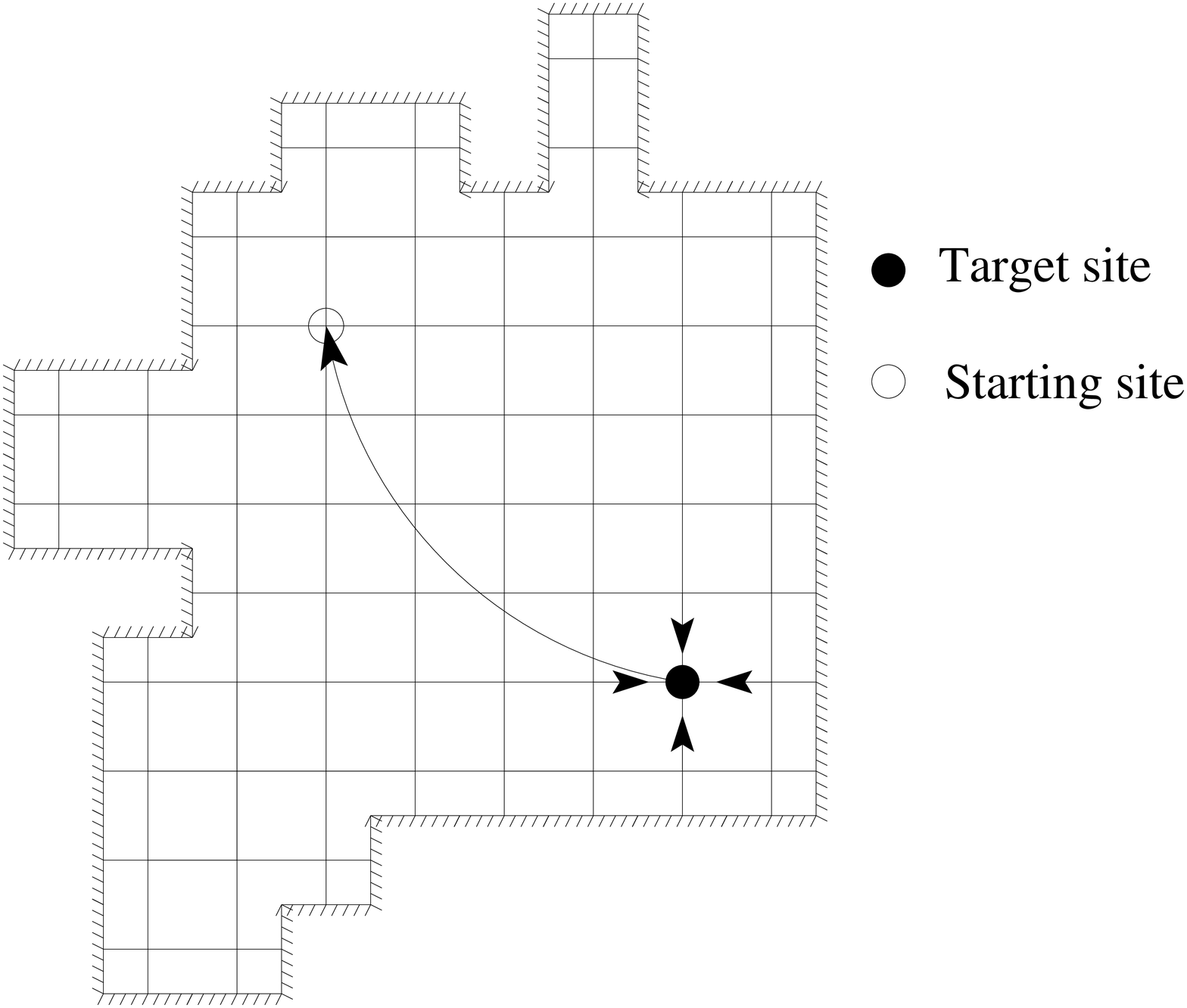}
\caption{Modifications of the original lattice: arrows denote one-way links.}
\label{astuce}
\end{figure}%

Kac's formula can be used to derive the MFPT $\Tm$ by slightly modifying the 
original lattice (see Fig.\ref{astuce}):  we suppress all the
original links starting from the target site $T$, and add a new
one-way link from $T$ to the starting point $S$, whereas all other links are 
unchanged. 
In this new lattice, any trajectory starting from $T$ goes to $S$ at its 
first
step, so that the MFRT to $T$ is just the MFPT from $S$ to $T$ in the
former lattice, plus one.

An exact, formal expression for the MFPT can thus be derived for 
the most general finite graph.
Consider $N$ points at positions ${\bf r}_1, \ldots, {\bf r}_N$ in an arbitrary
space. The transition rates from point $j$ to  
point $i$ are denoted $w_{ij}$. If we assume that one transition takes place
during each time unit we have: 
\begin{equation}
\sum_{i} w_{ij} = 1 
\end{equation}
Let ${\bf r}_T$ be the position of the target site,
${\bf r}_S$ be the  position of the starting site, and $\pi({\bf r})$ be the
stationary probability of the modified lattice. We write $\pi({\bf r}_T) = J$. 
According to Kac's formula, the MFRT to $T$ on the modified graph is 
$1/J$, so that the MFPT from $S$ to $T$ in the original graph is 
$ \langle \mathbf{T}\rangle
= 1/J - 1$. All we need to find is  the stationary
probability $\pi$. It  satisfies the following
equation:
\begin{equation}
\pi({\bf r}_i) = \sum_j w_{ij}\pi({\bf r}_j) +  J\delta_{iS} -
J w_{iT}
\end{equation}
where $\delta$ is the Kronecker symbol. 
To solve this equation, we define the auxiliary function
$\pi'$, such that $\pi'({\bf r}_i)=\pi({\bf r}_i) - J\delta_{iT}$.
It satisfies: 
\begin{equation}
\pi'({\bf r}_i) = \sum_j w_{ij}\pi'({\bf r}_j) +  J\delta_{iS} -
J \delta_{iT}
\label{eqpip}
\end{equation}
so that $\pi'$ has the following expression:
\begin{equation}
\pi'({\bf r}_i) = (1-J)\pi_0({\bf r}_i) + J H({\bf r}_i|{\bf r}_S) - 
J H({\bf r}_i|{\bf r}_T),
\label{valpip}
\end{equation}
where $\pi_0$ is the stationary probability of the original lattice, 
and $H$ is the discrete
pseudo-Green function \cite{Barton}, 
which satisfies the two following equations:
\begin{equation}
H({\bf r}_i|{\bf r}_j) = \sum_k  w_{ik} H({\bf r}_k|{\bf r}_j)+ 
\delta_{ij} - \frac1N
\label{pseudogreen}
\end{equation}
\begin{equation}
\sum_i H({\bf r}_i|{\bf r}_j) \equiv \bar{H},
\label{hbar}
\end{equation}
where $\bar{H}$ is independent of $j$. 
Moreover, if $w_{ij}$ is symmetric, which will be the case in all the 
practical cases considered, $H$ will also be symmetric in its 
arguments. The pseudo-Green function can be seen as a generalization of the 
usual infinite-space Green function to a bounded domain. Indeed, Eq. 
(\ref{pseudogreen}) without the $-1/N$ term corresponds to the definition 
of the infinite-space Green function, which would not have any solution for a 
finite domain with reflecting boundary conditions: it is necessary in this 
case to compensate the source term $\delta_{ij}$, and the simplest way to do so 
is to add the $-1/N$ term. The properties of this function are further 
discussed in Appendix \ref{AnnGreen}. 
We can thus see that the solution (\ref{valpip}) satisfies 
equation (\ref{eqpip}), and  ensures that $\pi$ is normalized.

The condition $\pi'({\bf r}_T)=0$ allows us to compute $J$ and to deduce
the following exact expression:
\begin{equation}
\langle \mathbf{T}\rangle = \frac{1}{\pi_0({\bf r}_T)} 
[ H({\bf r}_T|{\bf r}_T) - H({\bf r}_T|{\bf r}_S)  ]
\end{equation}
If $w_{ij}$ is symmetric, and we will consider that this is the case 
in the rest of the paper, we simply have $\pi_0 = 1/N$, and we get the simpler 
formula: 
\begin{equation}
\langle \mathbf{T}\rangle = N 
[ H({\bf r}_T|{\bf r}_T) - H({\bf r}_T|{\bf r}_S)  ]
\label{randwalk}
\end{equation}

\begin{figure}[t]
\centering\includegraphics[width = .7\linewidth,clip]{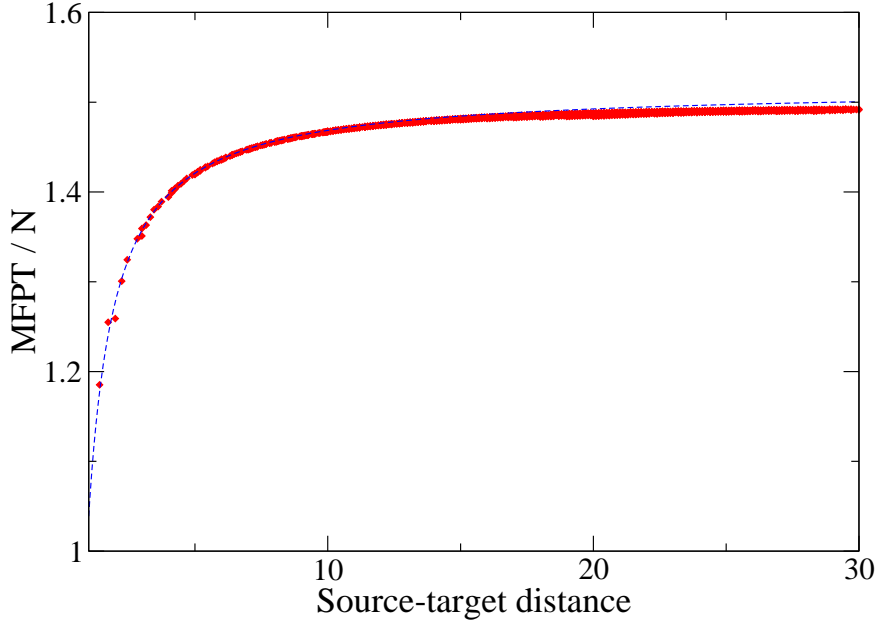}
\caption{(color online)
3D - Influence of the distance between the source and 
the target. Red crosses: simulations;  blue dashed line: evaluation
of the FPT with $H=G_0$.
The domain is a cube of side 41, the target being in the middle of it. 
All the simulation points correspond to different positions of the source.}
\label{distst}
\end{figure}

This result may be obtained by an alternative and complementary
approach. 
We consider that in the domain there is a constant flux $J$ of particles 
per time unit entering the domain at the source point $S$. The particles are 
absorbed when they reach the target, and, since all particles are eventually 
absorbed, we have an outcoming flux $J$ at the target. The average number
of particles in the domain satisfies: 
$\mathcal{N}=J\Tm $, which will allow the determination of $\Tm$. 
Indeed, the average density of particles $\rho({\bf r})$
satisfies the following equation: 
\begin{equation}
\rho({\bf r}_i) = \sum_j w_{ij}\rho({\bf r}_j) + J\delta_{iS} - J\delta_{iT}.
\label{alternative}
\end{equation}
The three terms of the equation correspond respectively to the diffusion
of particles, the incoming flux in $S$, and the outgoing flux in $T$. 
This is exactly the same equation as Eq.(\ref{eqpip}), with the same condition
$\rho({\bf r}_T) = 0$, and thus admits a similar solution, with the difference
that the total number of particles in the domain is not fixed a priori. 
The solution is thus: 
\begin{equation}
\rho({\bf r}_i) = \rho_0 + J H({\bf r}_i|{\bf r}_S) - J H({\bf
r}_i|{\bf r}_T)
\label{rhodiscret}
\end{equation}
which gives, with the condition $\rho({\bf r}_T) = 0$ and the 
relation $J\Tm = \mathcal{N} = N\rho_0$,
the same result as before for the mean first-passage time, namely 
Eq.(\ref{randwalk}).
This formula is equivalent to the one given in \cite{Noh}, but is
expressed in  terms of pseudo-Green functions. One advantage of the present
method is that it  may  be easily extended to more complex situations,
as it will be shown.   Another advantage is that, although the
pseudo-Green function $H$ is not known  in general, it is well suited
for approximations when the graph is a bounded regular 
lattice. The simplest one in this case is to approximate the
pseudo-Green function by its  infinite-space limit, the "usual" Green
function: $H({\bf r}|{\bf r}')\simeq  G_0({\bf r}-{\bf r}')$, which
satisfies:
\begin{equation}
G_0({\bf r}) = \frac{1}{\sigma}\sum_{{\bf r}' \in N({\bf r})}
G_0({\bf r}')+ \delta_{0{\bf r}}.
\label{green}
\end{equation} 
where $N({\bf r})$ is the ensemble of neighbours of ${\bf r}$, and $\sigma$ 
the coordination number of the lattice.  
The value of $G_0(0)$ and the asymptotic behaviour of $G_0$ are well-known
\cite{Hughes}. For instance, for the 3D cubic lattice, we have:  $G_0(0)
= 1.516386...$  and $G_0({\bf r}) \simeq 3/(2\pi r)$ for $r$ large.
For the 2D square lattice, we have  $G_0(0)-G_0({\bf r}) \simeq
(2/\pi) \ln(r) + (3/\pi) \ln2 +  2\gamma/\pi$, where
$\gamma$ is the Euler gamma constant, and 
$(3/\pi) \ln2 +  2\gamma/\pi = 1.029374...$ . These estimations of $G_0$ are
used for all the practical applications in the following. 
In some cases (especially in three dimensions when the target is far from 
any boundary), approximating $H$ by $G_0$ will give accurate results
(see Fig. \ref{distst}). The small correction is due to boundary effects, 
which are further discussed in Section \ref{discussion}. 
In other cases it will only give an order of magnitude.
In the case of a rectangular or parallepipedic
domain an exact expression of $H$ is known \cite{CondaminJCP}, and the 
FPT from any point to any other point in the domain can be computed exactly. 
This exact result and simple approximations, which can be used in other cases,
are given in Appendix \ref{approximations}.

\subsection{Application: absorbing opening in a reflecting boundary}

\begin{figure}[h!]%
\centering\includegraphics[width = .7\linewidth]{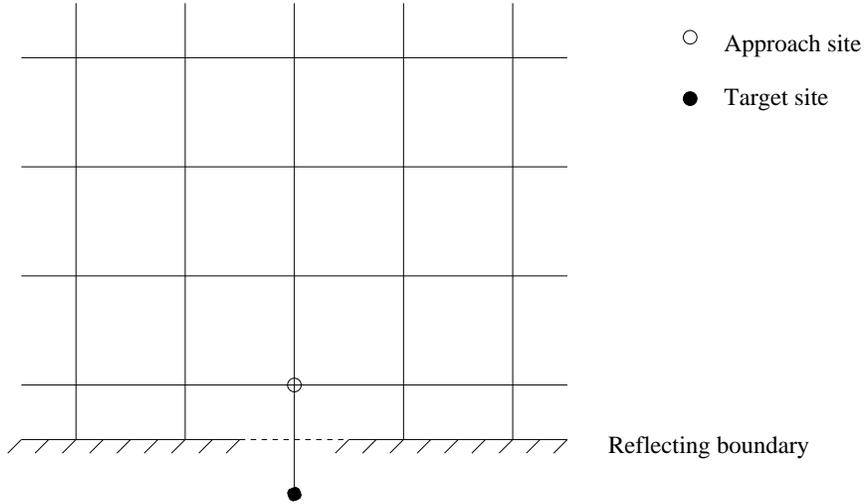}
\caption{Opening in a flat reflecting boundary}
\label{ouverture}
\end{figure}%

Another situation that may arise and can easily be dealt with  
is the case of an absorbing opening in a (locally) flat reflecting boundary 
of a bounded domain: we are 
interested in the mean time a particle takes to exit from the domain, 
if it may only exit by this opening. (see Fig. \ref{ouverture}). 
We only consider 
regular lattices of dimension $d=2$ or $3$. 
We can define a target site, just behind the flat boundary.
The problem here is that the pseudo-Green function for the domain plus the 
target site is difficult to compute, whereas the pseudo-Green function near
a flat boundary can be easily evaluated, and is even known exactly if 
the domain is rectangular or parallepipedic. To solve this problem, we will 
call the site next to the 
target the approach site $A$. We indeed have to go through this approach 
site in order to reach the target site. We will call 
$\langle\mathbf{T}\rangle_{ST}$ 
the average time to reach  the target site, starting from the source; 
$\langle\mathbf{T}\rangle_{SA}$ the average time to 
reach the approach site, still starting from the source; 
$\langle\mathbf{T}\rangle_{AA}$ the 
average time to return to the approach site, assuming the random walk 
does not go to the target site after exiting the approach site; 
$\langle\mathbf{T}\rangle_{AT}$ 
the average time to reach the target site, starting from the approach site. 
We have the following equations: 
\begin{equation}
\langle\mathbf{T}\rangle_{ST} = \langle\mathbf{T}\rangle_{SA} + 
\langle\mathbf{T}\rangle_{AT}
\end{equation}
since the random walk has to go through the approach site, and 
\begin{equation}
\langle\mathbf{T}\rangle_{AT} = \frac{2d-1}{2d}\left(\left< \mathbf{T}
\right>_{AA} +
\left<\mathbf{T}\right>_{AT}\right)+\frac1{2d}
\label{eqat}
\end{equation}
Once the random walker is at the approach site, it may either go directly 
to the target site (probability $\frac1{2d}$, where $d$ is the dimension
of the lattice) or go another way, in which case
it has to go back to the approach site before finding the target site.
Thus,
\begin{equation}
\left<\mathbf{T}\right>_{AT} = (2d-1)\left<\mathbf{T}\right>_{AA} + 1
\end{equation}
To compute $\langle \mathbf{T}\rangle_{AA}$, we have to remember that, if the boundary was fully 
reflecting, we would have the average return time: it is given by Kac's 
formula, and is $N$. We then have, with arguments similar to Eq. \ref{eqat}: 
\begin{equation}
N = \frac{2d-1}{2d}\langle \mathbf{T}\rangle_{AA}+ \frac1{2d}
\end{equation}
We then have: 
\begin{equation}
(2d-1) \langle \mathbf{T}\rangle_{AA}= 2dN - 1
\end{equation}
and thus:
\begin{equation}
\Tm_{AT}=2dN
\end{equation}
As for the average time needed to reach the approach site, starting from the 
starting site, it is exactly the same as in the case where the boundary is 
totally reflecting:
\begin{equation}
\langle \mathbf{T}\rangle_{SA} = N [ H({\bf r}_A|{\bf r}_A) - H({\bf
r}_A|{\bf r}_S)  ],
\end{equation}
and finally: 
\begin{equation}
\langle \mathbf{T}\rangle_{ST} = N [2d+ H({\bf r}_A|{\bf r}_A) - H({\bf
r}_A|{\bf r}_S)],
\end{equation}

To evaluate $H$, we have to take into account the effect
of the boundary. Since the boundary is flat, the simplest way to check the 
boundary condition is to write $H({\bf r}|{\bf r}') \simeq G_0({\bf r}-{\bf r}')+
G_0({\bf r}-s({\bf r}'))$, where $s({\bf r})$ is the point symmetrical to
${\bf r}$ with respect to the boundary. We will use this approximation in 
the following, (cf. Appendix 
\ref{approximations} for a discussion of this approximation)
We note $G_0(1)=G_0(0)-1$ the Green function for the 
sites surrounding the origin, and notice that $T$ is 
symmetrical to $A$ with respect to the boundary. 
The mean exit time is then: 
\begin{equation}
\langle\mathbf{T}\rangle_{ST} \simeq N[2d+G_0(0)+G_0(1)-
G_0({\bf r}_S-{\bf r}_A)-G_0({\bf r}_S-{\bf r}_T)]
\end{equation}

\subsection{Higher-order moments}

Moreover, we are able to evaluate the higher-order moments and distribution
of the FPT in the 3D case, provided the domain is not too elongated, i.e. 
the typical distance between a point and a boundary is $N^{1/3}$. 
The computation of the moments is detailed in Appendix 
\ref{computationdiscret}.
However, we cannot compute the higher-order moments and distribution of the 
FPT in two dimensions, or with a too elongated 3D domain. 
The computational reasons behind this are explained in Appendix 
\ref{computationdiscret}, but we will also explain it later from a physical 
point of view. 
We obtain the following result for higher-order moments: 
\begin{equation}
\left<\mathbf{T}^n\right>_i = n!N^n\left[ \left(
H({\bf r}_T|{\bf r}_T)-H({\bf r}_T|{\bf r}_i)\right)
\left( H({\bf r}_T|{\bf r}_T) - \bar{H} \right)^{n-1}
+\mathcal{O}(nN^{-2/3}) \right], 
\label{eqmoments}
\end{equation}
where $\bar{H}$ is defined by Eq. (\ref{hbar})

\begin{figure}[t]
\centering\includegraphics[width = .7\linewidth,clip]{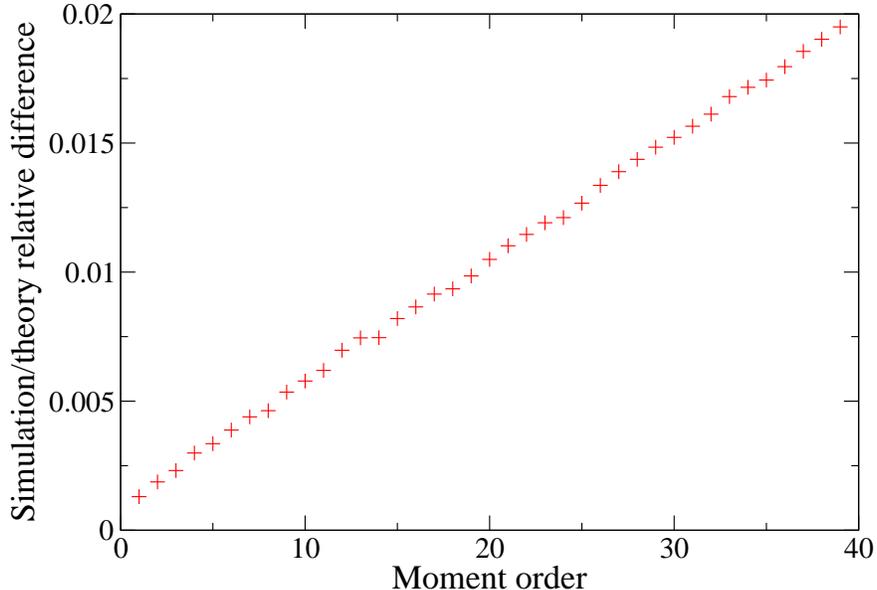}
\caption{(color online) 
3D - Relative difference between the simulations and the theoretical
prevision (\ref{eqmoments}). 
The domain is a cube of side 51 centered on the target at 
(0,0,0) and the source at (2,2,1). The order of magnitude of the relative 
difference is indeed $nN^{-2/3}$.}
\label{moments}
\end{figure}

To check these results, we computed the moments with a 
numerical simulation (cf. Appendix \ref{simulations} for the simulation 
method), and found (see Fig.\ref{moments}) a good agreement with the 
theoretical estimation (\ref{eqmoments}), where $H$ is approximated by $G_0$, 
and $\bar{H}$ is approximated by is its value
for a spherical domain, computed in the continuous limit, 
$\bar{H} = (18/5)(3/(4\pi))^{2/3}N^{-1/3}$ (cf. Eq. (\ref{valuehbar}) for the 
computation).

The study of the distribution in the limit of large $N$ will enable us to 
go even further. Indeed, if we neglect the corrections in $nN^{-2/3}$ in Eq.
(\ref{eqmoments}), the moments of $T/N$ are those of the following probability 
density $p$: 
\begin{equation}
p(t) = \left(\frac{H({\bf r}_T|{\bf r}_T)-H({\bf r}_T|{\bf r}_S)}
{(H({\bf r}_T|{\bf r}_T) - \bar{H})^2}\right)
\exp\left(-\frac{t}{H({\bf r}_T|{\bf r}_T) - \bar{H}}\right)
+ \frac{H({\bf r}_T|{\bf r}_S)-\bar{H}}{H({\bf r}_T|{\bf r}_T) - \bar{H}} 
\, \delta(t)
\label{probdist}
\end{equation}
The large-$N$ limit of this probability density is rigorous 
(since the corrections 
to the moments vanish). In this limit, 
$H({\bf r}_T|{\bf r}_T)$ tends to $G_0(0)$; $\bar{H}$ tends to $0$. 
Thus, the probability density of $\mathbf{T}/N$ tends to the following
probability density, the relative position of $i$ and $T$ being fixed:
\begin{equation}
p(t) = \left(\frac{G_0(0)-G_0({\bf r}_T-{\bf r}_S)}{G_0^2(0)}\right)
\exp\left(-\frac{t}{G_0(0)}\right)
+ \frac{G_0({\bf r}_T-{\bf r}_i)}{G_0(0)} \delta(t)
\label{problimit}
\end{equation}

\begin{figure}[t]
\centering\includegraphics[width = .7\linewidth,clip]{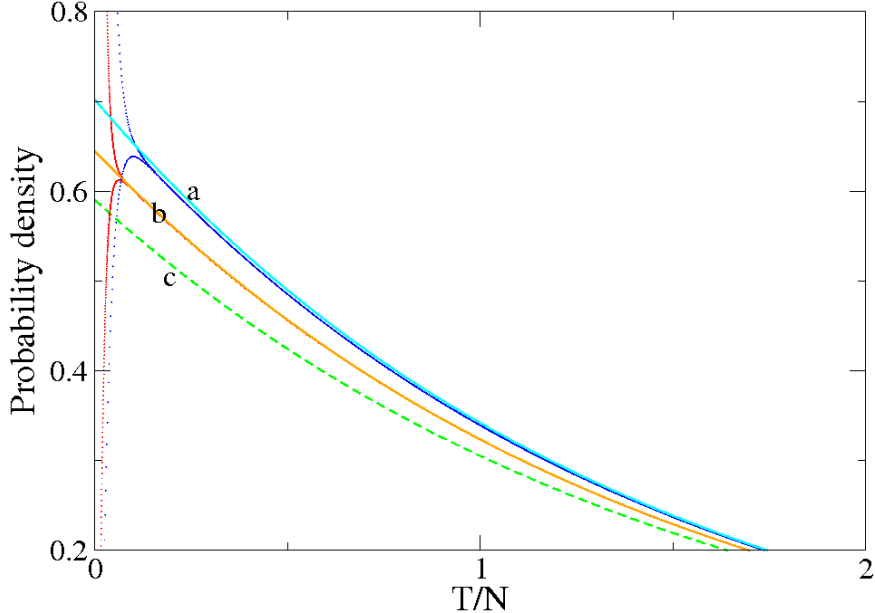}
\caption{(color online) 
Simulation of the probability distribution of the FPT, rescaled as
a probability density, with different 
domain sizes. In both cases, the target is in the middle of a cube at 
position (0,0,0) and the source is in (1,2,2). 
We plot the estimated density (\ref{probdist}) vs. numerical simulations
for different domain sizes. 
The dark blue (simulation) and cyan (estimated density) curves 
correspond to a cube of side 11 ($N = 1331$); The red (simulation) and 
orange (estimated density) correspond to a cube of side 21 ($N = 9261$)
; The green dashed curve 
corresponds to the high-$N$ limit (\ref{problimit}) of the probability 
density. For both domain sizes, the simulated distribution splits into two 
parts for short times and cannot be distinguished from the theoretical 
density afterwards. The labels a and b correspond to domain sizes of 
11 and 21, and the high-$N$ limit is labeled c. 
}
\label{distrib2}
\end{figure}

These results have been confronted to numerical simulations 
(Fig.\ref{distrib2}).
We computed the exact distribution for several domain sizes, 
and may notice that the curve divides in two at short times. 
This is 
due to the fact that, at short times, the parity of the step is important: as 
long as the walk does not touch the boundary, the distance between the starting
point and the walker has the same parity as the time elapsed. The time needed 
for the two curves to collapse into one shows very well the time needed to 
erase the memory of the starting position. 
The curves before this time correspond to the Dirac part of the 
probability density (\ref{probdist}); however, we can see that, 
once the two curves have collapsed, the resulting curves fit very well 
the theoretical prediction (\ref{probdist}), which is indeed more accurate than
the limit probability density (\ref{problimit}).

To analyse the physical meaning of this result, we may first notice that, if 
the probability density (\ref{probdist}) is averaged on the starting 
point, the Dirac part of the density vanishes, and we simply have 
an exponential distribution of the first-passage time. This 
property sheds a new light on the quasi-chemical approximation \cite{bere}, 
which assumes that if a particle starts randomly in a volume, and may only exit
through a small hole, it has a constant probability to exit at each time step. 
This approximation leads to an exponential distribution of the exit times. 
If we consider that the target site 
is the exit point for a particle, then the exit time is exactly the FPT. 
Thus, we have an evaluation of the accuracy of the quasi-chemical approximation
(or at least of its moments) in this case. 

The interpretation of the probability density (\ref{probdist})
is the following: 
the first part of the density, which decays exponentially, corresponds to
the decay of the probability distribution of the FPT if the particle starts 
randomly in the set of points. The second part corresponds to a particle 
reaching the target in a time negligible with respect to $N$. Here we must 
remember that a free 3D walk is transient: the particle may never reach the 
target in infinite space. Thus, one can interpret the Dirac term as the 
probability to reach the target \emph{without touching the boundaries}. For 
$N$ large enough it is equivalent to the probability to reach the target at 
all in infinite space. And, for this kind of trajectory, the probability 
distribution of the FPT does not depend of $N$, and, thus, the probability 
density of $\mathbf{T}/N$ will tend to $\delta(0)$ for large $N$. 
On the other hand, if the particle \emph{does} reach the boundary (it happens 
after a typical time $N^{2/3}$, since the boundaries are at a typical distance
$N^{1/3}$, and the typical time needed to cross a distance $r$ is $r^2$), 
its position will become random in a time 
negligible with respect to $N$, and, thus, the probability density of 
$\mathbf{T}/N$ will be the same as if the particle started in a random 
position in this latter case. This argument fails for an elongated domain,
which can be seen as a physical reason why we are not able to compute the FPT 
distribution in this case. 
We can check that the probability to reach the origin for a random
walk in infinite space is indeed $\frac{G_0({\bf r})}{G_0(0)}$ 
\cite{Hughes,Spitzer}.

\begin{figure}[t]
\centering\includegraphics[width = .7\linewidth,clip]{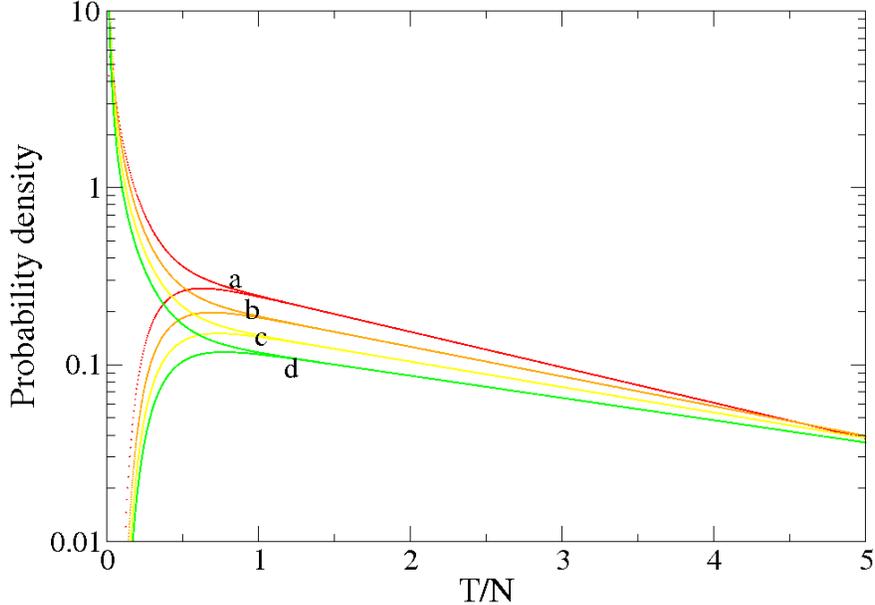}
\caption{(color online) 
Simulation of the distribution of the FPT in the 2D case, 
rescaled as a probability density.  
The domains are squares of different size;
The target is in the middle of the squares at position (0,0) and the 
source is in (2,3). The side of the squares are: 
21 (red curve [a]), 41 (orange curve [b]), 81 (yellow curve [c]) and 161 
(green curve [d]). 
The semi-logarithmic scale shows the long-time exponential decay. The 
splitting of the curves for short times is due to parity effects. 
}
\label{distrib2D}
\end{figure}

Here we can see an important physical difference between the 
2D and 3D cases: in two dimensions the random walk is recurrent. 
We can thus conclude that the large-$N$ limit probability density of 
$\mathbf{T}/N$
will be a simple delta function, since, in the limit of infinite space, the 
particle almost certainly reaches the target in a finite time, even if the 
MFPT is infinite! 
However, the probability distribution for $\emph{finite}$
$N$ will be much more difficult to compute: Indeed, there will also be two 
regimes, of low $\mathbf{T}/N$, when the particles have not touched the 
boundary and 
thus the distribution is the same as in free space; and the regime of 
high $\mathbf{T}/N$, where the distribution decays exponentially 
(since the system 
has lost the memory of its starting point). The transition between the 
two regimes happens at a finite $\mathbf{T}/N$ 
(since the time needed to reach the boundaries is of order $N$). 
Thus, the low $\mathbf{T}/N$ regime will have a much stronger influence on the 
values of the moments than on the $3D$ case, which may explain why the 
computation of the moments and distribution is much more delicate in this 
case. We can see in Fig. \ref{distrib2D} typical probability distributions for 
different domain sizes. One can very well see that the transition between 
the two regimes takes place at a finite $\mathbf{T}/N$ no matter the size 
of the domain, and that the long-time regime indeed corresponds to an 
exponential decay. 

\subsection{Case of two targets}

We can now assume that the lattice contains not one but two target points 
$T_1$ and $T_2$. 
The problems that may arise in this case are the mean time 
needed to reach one of the two targets, which we will call \emph{mean 
absorption time} and note $\langle\mathbf{T}\rangle$, and the 
\emph{splitting probabilities}, i.e. the 
probabilities $P_1$ to reach $T_1$ before $T_2$ and $P_2$ to reach $T_2$ 
before $T_1$. This model corresponds to the case of a diffusing particle 
which may be absorbed either by the target $T_1$ or the target $T_2$. 
We can also, even if it will be less straightforward,
 study the \emph{conditional} mean 
absorption time, i.e. the mean absorption time $\langle \mathbf{T}_1 \rangle$
(resp. $\langle \mathbf{T}_2 \rangle$), for particles which are absorbed by 
the target $T_1$ (resp. $T_2$).
This is relevant in many chemical applications \cite{Rice}, and may be useful
in biology to determine to which extent cellular variability may be controlled 
by diffusion \cite{Zon}.

To compute these quantities, it is more convenient to use the alternative 
approach presented on page \pageref{alternative}: we consider a constant 
incoming flux of particles $J$, and we 
have an average outcoming flux of particles $J_1$ in $T_1$, and $J_2$ in 
$T_2$. Since all particles are eventually absorbed, $J_1+J_2=J$. The probability
to reach the target $i$ is then $P_i = J_i/J$. 
The total number of particles $\mathcal{N}$ in the domain satisfies 
$\mathcal{N} = J\Tm$, and the mean density of particles satisfies the 
following equation: 
\begin{equation}
\rho({\bf r}_i) = \sum_j w_{ij}\rho({\bf r}_j) + J\delta_{iS} - J_1\delta_{iT_1}
- J_2\delta_{iT_2}
\end{equation}
We then get: 
\begin{equation}
\rho({\bf r}_i) = \rho_0 + J H({\bf r}_i|{\bf r}_S) - 
J_1 H({\bf r}_i|{\bf r}_{T_1}) - J_2 H({\bf r}_i|{\bf r}_{T_2}),
\label{2trho}
\end{equation}
then, writing $\rho({\bf r}_{T_1})=\rho({\bf r}_{T_2})=0$, we get the following
set of equations:
\begin{equation}
\left\{ \begin{array}{rcl}
\rho_0 + JH_{1s} - JP_1H_{01} - JP_2H_{12}
& = & 0 \\
\rho_0 + JH_{2s} - JP_2H_{02} - JP_1H_{12}
& = & 0 \\
P_1+P_2 & = & 1 \\
\end{array} \right.
\end{equation}
where $H_{12} = H({\bf r}_{T_1}|{\bf r}_{T_2})$ and, for $i = 1$ or $2$,  
$H_{is} = H({\bf r}_{T_i}|{\bf r}_S)$, 
$H_{0i} = H({\bf r}_{T_i}|{\bf r}_{T_i})$.
From this set of equation we can deduce $P_1$, $P_2$ and $\rho_0 = J\Tm/N$.
We thus get exact expressions for the mean absorption time 
and the splitting probabilities, respectively: 
\begin{equation}
\langle \mathbf{T}\rangle = N 
\frac{(H_{01}-H_{1s})(H_{02}-H_{2s}) - (H_{12}-H_{2s})(H_{12}-H_{1s})}
{H_{01}+H_{02}-2H_{12}}
\end{equation}
\begin{equation}
\left\{
\begin{array}{l}
P_1 = \frac{H_{1s}+H_{02}-H_{2s}-H_{12}}{H_{01}+H_{02}-2H_{12}} \\
P_2 = \frac{H_{2s}+H_{01}-H_{1s}-H_{12}}{H_{01}+H_{02}-2H_{12}} \\
\end{array}
\right.
\end{equation}

This result can be extended if necessary to more than two targets; if there are
$n$ targets, we have $n+1$ unknown variables ($\rho_0$ and the $n$ 
probabilities $P_k$), with $n+1$ equations, namely $\sum P_k = 1$ and the $n$ 
equations $\rho({\bf r}_{T_k})=0$, which is enough to determine all the 
unknown variables. However, this may quickly become computationally expensive
for a large number of targets.  

\begin{figure}[t]
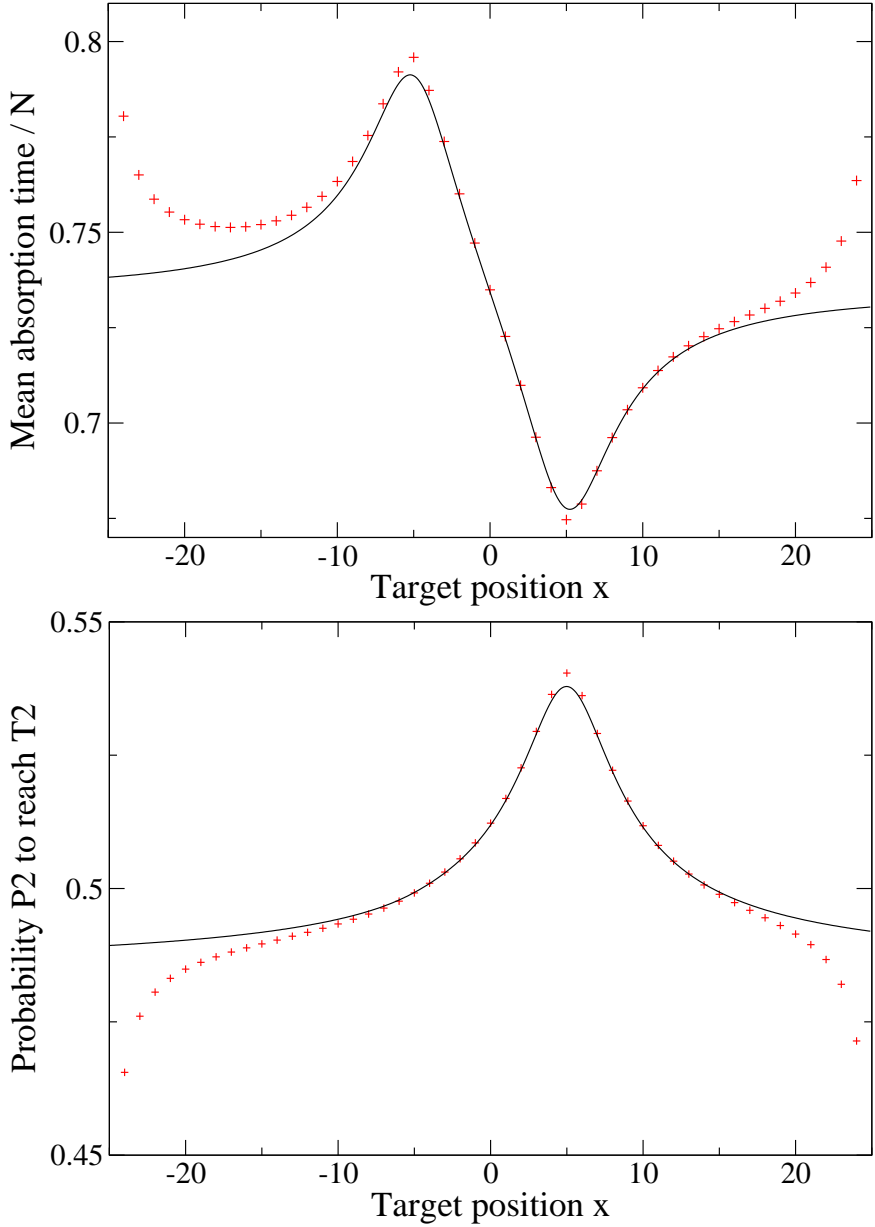

\begin{minipage}{.7\linewidth}
\centering\includegraphics[width = \linewidth,clip]{time2t}
\centering\includegraphics[width = \linewidth,clip]{prob2t}
\end{minipage}
\caption{(color online) 3D: two-target simulations.  Simulations (red crosses) 
vs. theory with the approximation $H=G_0$ (solid line). One target is fixed 
in (-5,0,0); the source is fixed 
in (5,0,0); the other target is at (x,3,0). The domain is a cube of
side 51, the middle is the point (0,0,0).} 
\label{2t}
\end{figure}

We compared the two-target results to simulations (Fig. \ref{2t}).
Note that if we use the exact value for $H$, which we can compute for a cube
(cf. Appendix  \ref{approximations}),
it is indeed impossible to see a difference between the theoretical predictions
and the simulations. 

It is interesting to underline an important qualitative difference between the
2D and 3D cases. In 3D, the furthest target always has a significant 
probability to be reached first,
since the most important terms in the probabilities $P_i$ are $H_{01}$ 
and $H_{02}$. In 2D, if a target is much closer from the source than the other,
it will almost certainly be reached first, 
since $H({\bf r}_i|{\bf r}_j)$
scales as $\ln|{\bf r}_i-{\bf r}_j|$. 
Actually, the probability for the furthest target to be reached first
decreases logarithmically. 
These properties are related to the transient character of the infinite 
3D walk, and the recurrent character of the 2D walk: indeed, an infinite 
2D walk explores all the sites of the lattice, whereas an infinite 3D walk 
does 
not; we may thus consider that the 2D walk will explore most of the sites 
surrounding the source before going much further, whereas the 3D walk will 
not, which qualitatively explains the difference of behaviour. 

\begin{figure}[t]
\centering\includegraphics[width = .7\linewidth,clip]{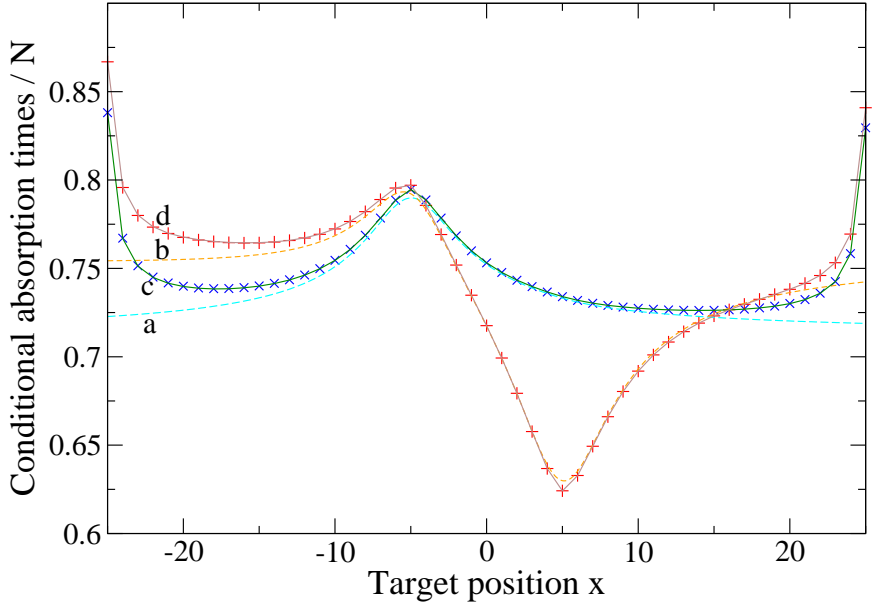}
\caption{3D: two-target simulations.  The conditions are identical to those 
of Fig.\ref{2t}; we show the conditional absorption times $\Tmm{1}$ 
(resp. $\Tmm{2}$). 
The blue crosses (resp. red plusses) show the results of the numerical 
simulations, the cyan [a] (resp.orange [b]) dashed line shows the 
theoretical expression 
(\ref{condtimes}) with $H=G_0$, the green [c] (resp.brown [d]) solid line 
shows the theoretical expression with the exact value of $H$ 
(\ref{exactcubic}).
}
\label{time2t}
\end{figure}

We can also determine the conditional absorption times $\Tmm{1}$ and 
$\Tmm{2}$. For this, we will compute $\mathcal{N}_k$, the average number of 
particles in the domain which will eventually be absorbed by $T_k$.
We have $\mathcal{N}_k = J_k\Tmm{k}$, which will allow us to compute $\Tmm{k}$.
To compute $\mathcal{N}_k$, we can simply notice that the density of particles 
that will eventually be absorbed by $T_k$ at the point $i$ is simply 
$\rho({\bf r}_i)P_k({\bf r}_i)$, where $P_k({\bf r}_i)$ is the probability to 
be absorbed by $T_k$ if the walk starts from $i$. 
We thus have: 
\begin{equation}
\mathcal{N}_k = \sum_{i}\rho({\bf r}_i)P_k({\bf r}_i)
\label{Ncondtimes}
\end{equation}
This equation is exact but may prove quite difficult to compute, especially 
in two dimensions if $H$ is not known exactly. 
However, in three dimensions, we may use the same kind of approximations 
as for the computation of the high-order moments of the FPT 
(with the same limitations, i.e. the 3D domain should not be too elongated)
to estimate 
the conditional probabilities. 
If we note $H_{iS} = H({\bf r}_i|{\bf r}_s)$ and 
$H_{ik}=H({\bf r}_i|{\bf r}_{T_k})$, we have: 
\begin{equation}
\mathcal{N}_1 = \sum_i\frac{(H_{i1}-H_{i2}+H_{02}-H_{12})
(\rho_0+JH_{iS}-J_1H_{i1}-J_2H_{i2})}{H_{01}+H_{02}-2H_{12}}
\end{equation}
We use the properties 
$\sum_i H({\bf r}_i|{\bf r}_j) = N\bar{H}$ (cf. Eq. (\ref{hbar}))
and
$\sum_i H({\bf r}_i|{\bf r}_j)H({\bf r}_i|{\bf r}_k) = \mathcal{O}(N^{1/3})$
(cf. Eq. \ref{scaling})
to write: 
\begin{equation}
\mathcal{N}_1 = N\frac{(H_{02}-H_{12})\rho_0 + \mathcal{O}(N^{-2/3})}{H_{01}+
H_{02}-2H_{12}}
\end{equation}
And we can conclude: 
\begin{equation}
\Tmm{1} = \frac{1}{P_1}\frac{H_{02}-H_{12}+\mathcal{O}(N^{-2/3})}{H_{01}+
H_{02}-2H_{12}}\Tm
\label{condtimes}
\end{equation}
The expression for $\Tmm{2}$ is of course equivalent. 
This expression is not exact, but is very accurate: the relative 
difference between the numerical simulations (see Fig.\ref{time2t}) and
the expression (\ref{condtimes}) is of about $0.01 \%$, for a domain of size
$N = 51^3$. 

Finally, we have a wide range of quantities which can be computed exactly, 
or with a very good accuracy, provided we know the pseudo-Green function $H$. 
Unfortunately, there are only a few cases in which it can be computed exactly.
Otherwise, we will have to use approximations, which, of course, give less 
accurate results. Both exact results and approximations are detailed in 
Appendix
\ref{approximations}.

\section{Brownian motion on continuous media\label{continu}}

\begin{figure}[t]
\centering\includegraphics[width = .4\linewidth,clip]{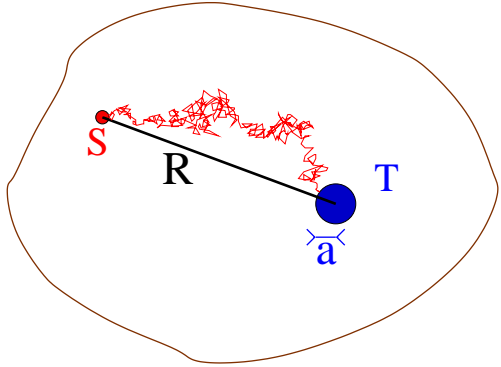}
\caption{(color online) Continuous problem}
\label{schemabrown}
\end{figure}

We may consider a similar problem in a continuous medium (see Fig. 
\ref{schemabrown}): if we 
have a Brownian
motion whose diffusion coefficient is $D$, how much time does it take to reach
a target? A difference with the discrete case is that the target has a 
finite size $a$ which is an important parameter of the problem. We will 
consider a spherical target $T$, of radius $a$, centered in ${\bf r}_T$. 
The Brownian motion starts from the starting point $S$ (its position is denoted
by ${\bf r}_S$). It is restricted to a domain 
$\mathcal{D}$ of volume $V$ (for $2D$ domains we will call the 
area $A$) , and we note $\mathcal{D}^*$ the domain deprived 
of the target. 
We will derive the same quantities as in the discrete case, but the 
results are this time only approximate; we can thus add some refinements to 
the method, in order to increase the accuracy. These refinements are given
in Appendix \ref{refinements}, and are used 
in practical computations of the MFPT in Appendix \ref{approximations}
 when the target is close to a boundary. It should be 
emphasised that, in the cases where the pseudo-Green function is known, such 
as the case of a spherical domain, the method gives accurate explicit 
expressions for all the MFPT and the other quantities studied here.

\subsection{Mean first passage time}

The mean first passage time (MFPT) $\Tmr{s}$ at the target satisfies the 
following equations \cite{Risken}:
\begin{equation}
 D \Delta \Tmr{s}=-1\;{\rm if}\; {\bf r_s}\in{\mathcal D}^*
\end{equation}
\begin{equation}
  \Tmr{s}=0 \;{\rm if}\; {\bf r_s}\in \Sigma_{\rm abs}
\end{equation}
\begin{equation}
  \partial_n\Tmr{s}=0 \;{\rm if}\; {\bf r_s}\in \Sigma_{\rm refl}
\end{equation}
where $\Sigma_{\rm abs}$ (resp. $\Sigma_{\rm refl}$) stands for the surface of the absorbing
target sphere (resp. the reflecting confining surface) and $\partial_n$ 
denotes the normal derivative. The boundaries have to be regular enough (twice 
continuously differentiable is sufficient, but not necessary) for these 
definitions to make sense. 

To solve this problem, we introduce the following Green function $G({\bf r}|{\bf r'})$ defined by
\begin{equation}\label{G1}
 -\Delta  G({\bf r}|{\bf r'}) =\delta({\bf r}-{\bf r'}) \;{\rm if}\; {\bf r}\in{\mathcal D}^*
\end{equation}
\begin{equation}\label{G2}
 G({\bf r}|{\bf r'})=0 \;{\rm if}\; {\bf r}\in \Sigma_{\rm abs}
\end{equation}
\begin{equation}\label{G3}
  \partial_n G({\bf r}|{\bf r'})=0 \;{\rm if}\; {\bf r}\in \Sigma_{\rm refl}
\end{equation}
Note that this Green function may also be seen as the stationary density 
of particles if there is an unit incoming flux of particles in ${\bf r}'$, 
and the diffusion coefficient is set to $1$. It should not be confused with 
the free Green function $G_0$, and is rather the continuous equivalent of the 
average density of particles $\rho$ defined in Eq. (\ref{alternative}) with 
$J=1$. It depends implicitly on the target 
position through Eq. (\ref{G2}). 

Using Green's formula, 
\begin{equation}
 \int_{{\mathcal D}^*}\left(\left< \mathbf{T(r)}\right>\Delta G({\bf r}|{\bf r'})
-G({\bf r}|{\bf r'})\Delta\left< \mathbf{T(r)}\right>\right)d^d{\bf r}=
\int_{\Sigma_{\rm abs}+\Sigma_{\rm refl}} \left(\left< \mathbf{T(r)}\right>
\partial_n G({\bf r}|{\bf r'})-G({\bf r}|{\bf r'})
\partial_n\left< \mathbf{T(r)}\right>\right)d^{d-1}{\bf r},
\end{equation}
we easily find that the MFPT is given by
\begin{equation}\label{Texpl}
\Tmr{S}=\frac{1}{D}\int_{\mathcal{D}^*} G({\bf r}|{\bf r}_S) 
d^d{\bf r}
\end{equation}
 
To approximate $G({\bf r}|{\bf r}_S)$ we can use a  
direct transposition to the continuous case of Eq. 
(\ref{rhodiscret}) :
\begin{equation}\label{Gapprox}
 G({\bf r}|{\bf r}_S)\simeq \rho_0({\bf r}_S)+H({\bf r}|{\bf r}_S)-
H({\bf r}|{\bf r}_T)
\end{equation}  
where $\rho_0$ is defined by $G({\bf r}|{\bf
r}_S)\simeq0$  if ${\bf r}\in \Sigma_{\rm abs}$ and  $H({\bf r}|{\bf
r}')$ is the pseudo-Green function \cite{Barton}, which  satisfies:
\begin{equation}\label{1} 
- \Delta  H({\bf r}|{\bf r'}) =\delta({\bf
r}-{\bf r'}) -\frac{1}{V}\;{\rm if}\; {\bf r}\in{\cal D}
\end{equation}
\begin{equation}\label{2}
  \partial_n H({\bf r}|{\bf r'})=0 \;{\rm if}\; {\bf r}\in \Sigma_{\rm
refl}
\end{equation}
\begin{equation}\label{3} 
H({\bf r}|{\bf r'})=H({\bf r'}|{\bf r})
\end{equation}
\begin{equation}\label{4} 
\int_{\mathcal{D}}H({\bf r}'|{\bf r})d^d{\bf
r}' \equiv V\bar{H},
\end{equation} 
$\bar{H}$ being independent of ${\bf r}$. This latter
equation can be easily  deduced from the three previous ones.

The choice (\ref{Gapprox}) of $G({\bf r}|{\bf r'})$ is the simplest one which
satisfies formally Eqs (\ref{G1}) and (\ref{G3}). However,
(\ref{G2}) can only be approximately satisfied. To take into account
this latter equation, we will approximate, on the target sphere,
$H({\bf r}|{\bf r}_S)$ by $H({\bf r}_T|{\bf r}_S)$ and $H({\bf r}|{\bf r}_T)$
by $G_0({\bf r}-{\bf r}_T)+H^*({\bf r}_T|{\bf r}_T)$, where $G_0$ is the 
well-known free Green function( $(2\pi)^{-1} \ln(r)$ in 
2D, $1/(4\pi r)$ in 3D), and $H^*$ is defined by:
\begin{equation}
H^*({\bf r}|{\bf r}') \equiv H({\bf r}|{\bf r}') - G_0({\bf r}-{\bf r}'). 
\end{equation}
Note that $H^*({\bf r}|{\bf r}_T)$ has no singularity in ${\bf r}_T$. 
Thus on the surface of the target sphere we have: 
\begin{equation}
\rho_0({\bf r}_S)+H({\bf r}_T|{\bf r}_S)-G_0(a)-H^*({\bf r}_T|{\bf r}_T) = 0 ,
\end{equation}
where $G_0(a)$ is the value of $G_0({\bf r})$ when $|{\bf r}| = a$. 
We can now compute 
\begin{equation}
\Tmr{S} = \frac1D \int_{\mathcal{D}^*} \left( \rho_0({\bf r}_S) + 
H({\bf r}|{\bf r}_S) - H({\bf r}|{\bf r}_T) \right) d^d{\bf r}
\end{equation}
Since the target is small compared to the domain, the integral over 
$\mathcal{D}^*$ is almost equal to the integral over $\mathcal{D}$, the 
relative order of magnitude of the correction being $a^3/V$ in 3D and 
$a^2/A$ in 2D. 
Using the property (\ref{4}), we can then compute the integral, 
and find the result: 
\begin{equation}
\Tmr{S} = \frac{V\rho_0({\bf r}_S)}{D} = \frac{V}{D}\left(G_0(a)+
H^*({\bf r}_T|{\bf r}_T)-H({\bf r}_T|{\bf r}_S)\right)+ \mathcal{O}
\left(\frac{a^dG_0(a)}{D}\right)
\label{simplebrownian}
\end{equation}
This equation is very close to (\ref{randwalk}), with the correspondence 
$H({\bf r}|{\bf r}) \rightarrow G_0(a)+H^*({\bf r}|{\bf r})$, but one should
pay attention to the fact that this is only an approximation! One may expect 
deviations from this expression when the variations of $H({\bf r}|{\bf r}_S)$ 
or $H^*({\bf r}|{\bf r}_T)$ will not be negligible over the target sphere; 
it corresponds to the cases when the target is either near the source or 
near a boundary. 
However, if we compare the expression obtained with 
simulations (see Fig.\ref{ctpos2D}) when the target is near the source, we
see no such deviation; this is justified in Appendix \ref{refinements}. 
On the other hand, there is indeed a deviation near the boundaries. This 
deviation scales as $a/d$ in two dimensions, or $a/d^2$ in three dimensions, 
where $d$ is the distance between the target and the boundary. It is possible
to compute a correction, which is explicited in Appendix \ref{refinements}, 
and used in practical applications in 
Appendix \ref{approximations}.

\begin{figure}[t]
\centering\includegraphics[width = .7\linewidth,clip]{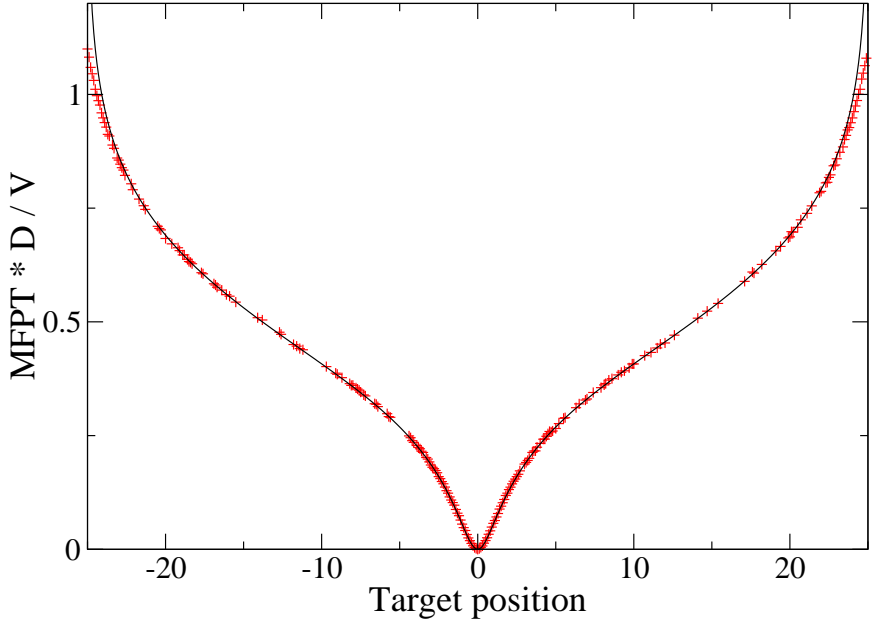}
\caption{(color online)
Brownian motion on a 2D disk of radius 25 centered on (0,0); 
the source is in (0,1) and the target of radius 1 is on (x,0). Red crosses: 
simulations; black solid line: estimation (\ref{simplebrownian}) with the
exact function $H$ for a sphere given by Eq. (\ref{exactdisk})} 
\label{ctpos2D}
\end{figure}

The exact value of $H$ is known analytically for disks and spheres 
\cite{Barton}, we will detail this in Appendix \ref{approximations}. This is 
why we will test the expressions we obtain in such geometries. 
If no exact expression is known, the simplest approximation of $H$ is simply
$H = G_0$. More accurate approximations are also discussed in Appendix
\ref{approximations}. 
We give the estimations of $\Tmr{S}$ with the basic approximation, to show 
the order of magnitude: 
\begin{equation}
\Tmr{S} = \frac{V}{4\pi D}\left(\frac1a-\frac1R\right) \;\;{\rm (3D)}
\label{base3D}
\end{equation}
\begin{equation}
\Tmr{S} = \frac{A}{2\pi D}\ln\frac{R}{a} \;\;{\rm (2D)}
\label{base2D}
\end{equation}
$R$ being the source-target distance. 
This already improves the (exact) asymptotic results of Pinsky 
\cite{Pinsky}, which only give the leading term in $a$.

\subsection{Higher-order moments}

The higher-order moments and density of the FPT in the three-dimensional
case can also be computed. The computation is detailed in Appendix 
\ref{computationcontinu}; the results are quite similar to the results 
obtained in the discrete case, and the physical interpretation is essentially
the same. The results obtained are the following: 

\begin{equation}
\Tmur{n}{S} = \frac{n!V^n}{D^n}\left[
\left(G_0(a) + H^*({\bf r}_T|{\bf r}_T) - H({\bf r}_T|{\bf r}_S)\right)
\left(G_0(a) + H^*({\bf r}_T|{\bf r}_T) - \bar{H}\right)^{n-1}
+\mathcal{O}\left(nV^{-2/3}a^{2-n} \right) \right]
\end{equation}
We may also deduce from this information about the probability density
of the absorption time $p(t)$: 
If we drop the term $\mathcal{O}\left(nV^{-2/3}a^{2-n}\right)$,
we have: 
\begin{eqnarray}
p(t) & = & \frac{D}{V}
\frac{G_0(a)+H^*({\bf r}_T|{\bf r}_T )-H({\bf r}_T|{\bf r}_S)}{\left(G_0(a)+H^*({\bf r}_T|{\bf r}_T )-\bar{H}\right)^2}
\exp\left(\frac{-Dt}{V\left(G_0(a)+H^*({\bf r}_T|{\bf r}_T )-\bar{H}\right)}\right) \\
&&
+ \frac{H({\bf r}_T|{\bf r}_S)-\bar{H}}{G_0(a)+H^*({\bf r}_T|{\bf r}_T )-\bar{H}} \delta(t) \nonumber
\end{eqnarray}
In the limit $a \rightarrow 0$, with the position of ${\bf r}_S$ fixed, 
the $H$ terms are constant since they only depend on the shape of the 
domain, and $G_0(a)$ tends towards infinity.
The probability density then simply becomes exponential: 
\begin{equation}
p(t) = \frac{4\pi aD}{V}\exp\left(-\frac{4\pi aDt}{V}\right)
\end{equation}
In the limit $a \rightarrow 0$, with the quantity $R/a$ fixed, 
$H({\bf r}_S|{\bf r}_T) \sim G_0(R)$, 
and the probability density becomes: 
\begin{equation}
p(t) = 
\frac{4\pi Da}{V}\left(1 -\frac{a}R\right)
\exp\left(-\frac{4\pi aDt}{V}\right)
+ \frac{a}{R} \delta(t)
\end{equation}

We did not test these results with a numerical simulation, since the 
continuous simulation method (see Appendix \ref{simulations}) is not 
adapted to the computation of the FPT density, and would require a 
large computation time to give accurate results. Furthermore, the 
approximations made (cf. Appendix \ref{computation}) are the same as on the 
discrete case, and the discrete results have been successfully compared to 
an exact numerical simulation (cf. Fig. \ref{distrib2}).

\subsection{Case of two targets}

For the case of two targets, we will compute the same quantities as in the 
discrete case; however, we may notice that the radius $a_1$ and $a_2$
 of the two targets 
may differ, which adds another parameter to the problem. 
With two targets, we will use the same Green function as before, but 
$\Sigma_{\rm abs}=\Sigma_1+\Sigma_2$ will be the reunion of the surfaces of 
the two absorbing target spheres. 
The mean absorption time $\Tmr{S}$ satisfies the equation (\ref{Texpl}); 
the splitting probability $P_1({\bf r}_S)$  satisfies the 
following equations \cite{VanKampen}: 
\begin{equation}
\Delta P_1({\bf r}) = 0
\end{equation}
\begin{equation}
P_1({\bf r}) = 1 \; {\rm if} \; {\bf r} \in \Sigma_1
\end{equation}
\begin{equation}
P_1({\bf r}) = 0 \; {\rm if} \; {\bf r} \in \Sigma_2
\end{equation}
\begin{equation}
\partial_n P_1({\bf r}) = 0 \; {\rm if} \; {\bf r} \in \Sigma_{\rm refl}
\end{equation}
Using Green's formula, we get: 
\begin{equation}
P_1({\bf r}_S) = - \int_{\Sigma_1} \partial_n G({\bf r}|{\bf r}_S) d{\bf r}
\label{expP1},
\end{equation}
The expression for $P_2$ is of course similar. 
Note that the normal derivative is oriented towards the inside of the target.
A simple approximation of $G$, equivalent to the discrete Eq. (\ref{2trho}) is:
\begin{equation}
G({\bf r}|{\bf r}_S) = \rho_0({\bf r}_S) + H({\bf r}|{\bf r}_S) - 
P_1({\bf r}_S)H({\bf r}|{\bf r}_{T_1}) - P_2({\bf r}_S) H({\bf r}|{\bf r}_{T_2}).
\end{equation}
This expression satisfies Eq.(\ref{G1}),(\ref{G3}) and (\ref{expP1}), and
$\rho_0$, $P_1$ and $P_2$ are set in order to satisfy Eq.(\ref{G2}) 
approximately. We use the same approximations as in the one-target case, 
which gives the following set of equations: 
\begin{equation}
\left\{ \begin{array}{rcl}
\rho_0({\bf r}_S) + H_{1s} - P_1H_{01} - P_2H_{12}
& = & 0 \\
\rho_0({\bf r}_S) + H_{2s} - P_2H_{02} - P_1H_{12}
& = & 0 \\
P_1+P_2 & = & 1 \\
\end{array} \right.
\label{2targetarray}
\end{equation}
where $H_{12} = H({\bf r}_{T_1}|{\bf r}_{T_2})$ and, for $i = 1$ or $2$,  
$H_{is} = H({\bf r}_{T_i}|{\bf r}_S)$, 
$H_{0i} = G_0(a_i) + H^*({\bf r}_{T_i}|{\bf r}_{T_i})$.
These equations are exactly identical to the discrete equations, only 
the meaning of the $H_{0i}$ changes.
We thus can deduce, using the same relation between $\rho_0$ and $\Tm$ as
in Eq. (\ref{simplebrownian}):
\begin{equation}
\Tmr{S} = \frac{V}{D}
\frac{(H_{01}-H_{1s})(H_{02}-H_{2s}) - (H_{12}-H_{2s})(H_{12}-H_{1s})}
{H_{01}+H_{02}-2H_{12}}
\label{tcontinu}
\end{equation}
\begin{equation}
\left\{
\begin{array}{l}
P_1 = \frac{H_{1s}+H_{02}-H_{2s}-H_{12}}{H_{01}+H_{02}-2H_{12}} \\
P_2 = \frac{H_{2s}+H_{01}-H_{1s}-H_{12}}{H_{01}+H_{02}-2H_{12}} \\
\end{array}
\right.
\label{pcontinu}
\end{equation}

\begin{figure}[t]
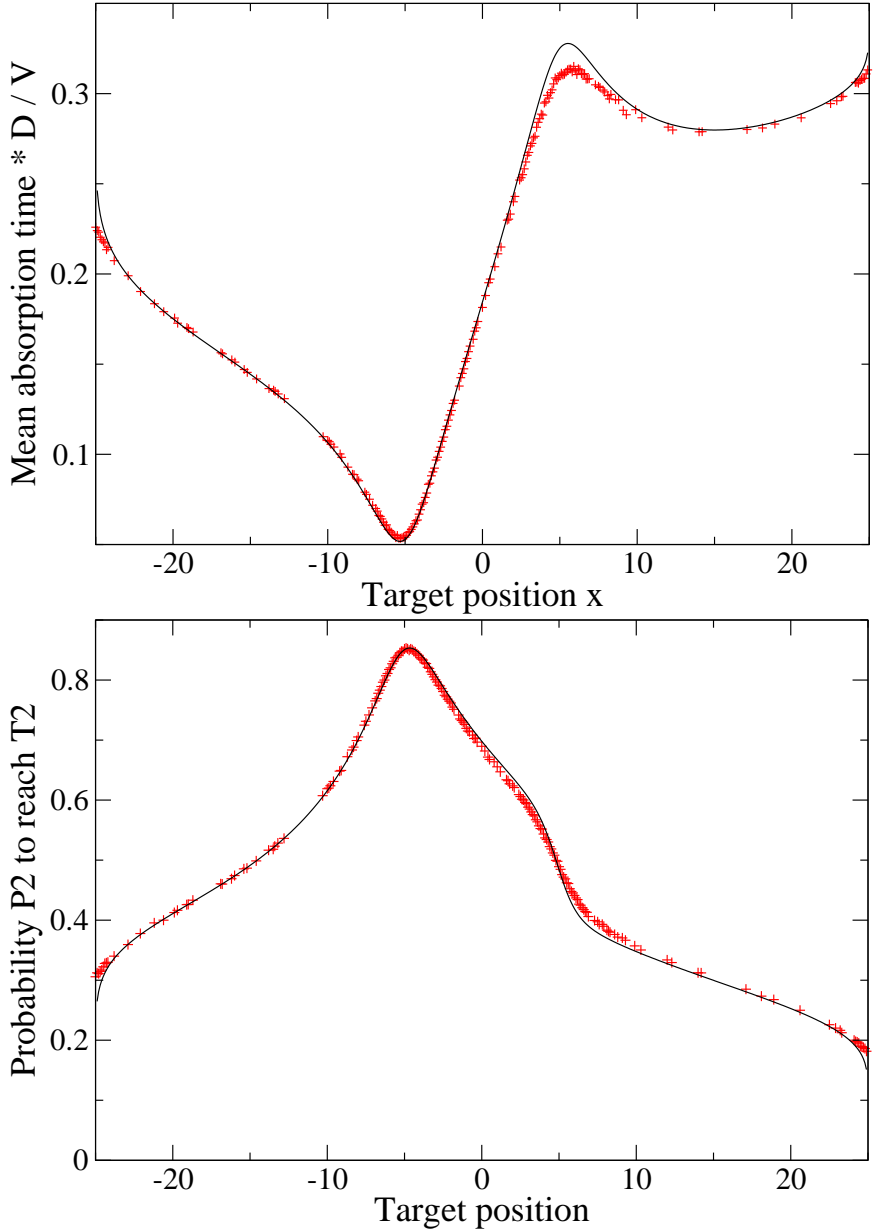

\begin{minipage}{.7\linewidth}
\centering\includegraphics[width = \linewidth,clip]{ctime2D}
\centering\includegraphics[width = \linewidth,clip]{cprob2D}
\end{minipage}
\caption{(color online) 
Brownian motion on a 2D disk of radius 25 centered on (0,0); D=1; 
the source is in (-5,2) and the two targets of radius 1 are on (5,2) ($T_1$)
and (x,0) ($T_2$). Red crosses: 
simulations; black solid line: estimations (\ref{tcontinu}) and 
(\ref{pcontinu}) with the exact function $H$ for a disk (\ref{exactdisk}).} 
\label{c2t2D}
\end{figure}

\begin{figure}[t]
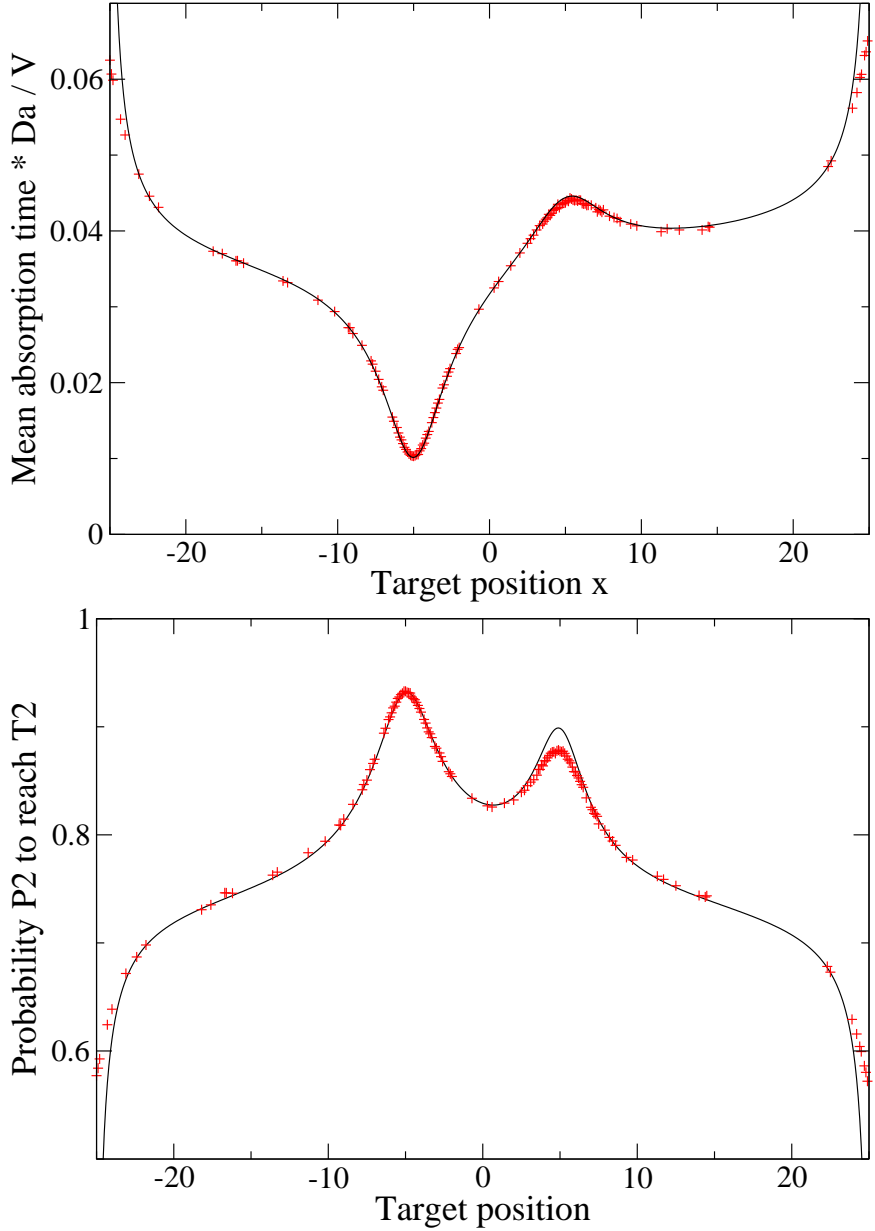

\begin{minipage}{.7\linewidth}
\centering\includegraphics[width = \linewidth,clip]{ctime3D}
\centering\includegraphics[width = \linewidth,clip]{cprob3D}
\end{minipage}
\caption{(color online) 
Brownian motion on a 3D sphere of radius 25 centered on (0,0,0); D=1; 
the source is in (-5,2,0) and the two targets are on (5,2,0) ($T_1$, of 
radius $0.5$) and (x,0,0) ($T_2$, of radius $1.5$). Red crosses: 
simulations; black solid line: estimations (\ref{tcontinu}), 
(\ref{pcontinu}), with the exact function $H$ for a sphere (\ref{exactsphere}).
} 
\label{c2t3D}
\end{figure}

We show in Fig. \ref{c2t2D} and \ref{c2t3D} the results of the numerical 
simulations. 
We can see that they are accurate, with a small correction (the relative 
correction scales as 
$(a/d)\ln(d/a)$ in 2D or $a^2/d^2$ in 3D, $d$ being the distance between the 
two targets) when the two targets are close (an explicit correction is given 
in Appendix \ref{refinements}), and when one target is 
near a boundary (exactly as in the one-target case). 

The curves themselves deserve a few qualitative remarks. Unsurprisingly, the
splitting probability $P_2$ is maximal when $T_2$ is the closest to the source. 
When the two targets have different sizes, an interesting phenomenon appears
(Fig. \ref{c2t3D}): the probability to hit the largest target ($T_2$) has a 
second maximum when it is close to the other target. One can understand 
this by a scaling argument. If the two targets are far away, $P_1$ will be 
about $a_1/(a_1+a_2)$. If the two targets touch one another, and $a_1 \ll a_2$, 
then the target $T_1$ covers a surface $\pi a_1^2$ of the target $T_2$. It 
can thus be expected that the probability $P_1$ will scale as $a_1^2/a_2^2$, 
and thus be much lower than if the two targets were far away. These arguments
are for the 3D case, but the qualitative behavior would be the same in the 
2D case. However, the behavior of the splitting probabilities when one target
is much further than the other from the source will be different in 2D and 
3D, for the same reasons as in the discrete case. In the figures the domain is
not large enough to make the difference obvious. 
The mean absorption time has a similar qualitative behavior in both cases: 
an unsurprising minimum when the moving target is close to the source, maxima 
when the moving target is near a boundary, due to boundary effects, and 
a maximum when the two targets are close, which deserves a few more comments. 
This could indeed be predicted directly from Eq.(\ref{tcontinu}), but, 
physically, this comes from the fact that, if the two targets are close, 
a particle undergoing a Brownian motion, which reaches one target, often would
have reached the other shortly afterwards in a single target situation. Thus,
the mean time gained, compared to the single target situation, will be 
much lower when the two targets are close. To analyse the values themselves, 
one should keep in mind that the times are normalized by $V/D$; the order of 
magnitude of the normalized times will then be $G_0(a)-G_0(R)$, which explains 
the values around $0.05$ obtained in the 3D case.

As for the conditional FPTs $\Tmmr{1}{S}$ and $\Tmmr{2}{S}$, 
we have the following relations \cite{VanKampen}: 
\begin{equation}
 D \Delta (P_1({\bf r})\Tmmr{1}{})= -P_1({\bf r}) \;{\rm if}\; {\bf r}
\in{\mathcal D}^*
\end{equation}
\begin{equation}
 P_1({\bf r})\Tmmr{1}{} = 0 \;{\rm if}\; {\bf r}\in \Sigma_{\rm abs}
\end{equation}
\begin{equation}
  \partial_n(P_1({\bf r})\Tmmr{1}{})=0 \;{\rm if}\; {\bf r}\in \Sigma_{\rm refl},
\end{equation}
and of course the equivalent relations for $\Tmmr{2}{}$. 
We use as usual Green's formula, and obtain: 
\begin{equation}
P_1({\bf r}_S)\Tmmr{1}{S} = \int_{\mathcal{D}^*} G({\bf r}|{\bf r}_S)P_1({\bf r})
d{\bf r}
\end{equation}
This equation is very similar to the discrete Eq. (\ref{Ncondtimes}) and the 
following calculations for the 3D case are exactly identical, and give: 
\begin{equation}
\Tmmr{1}{S} = \frac{1}{P_1({\bf r}_S)}
\frac{H_{02}-H_{12}+\mathcal{O}(aV^{-2/3})}{H_{01}+H_{02}-2H_{12}}\Tmr{S}
\label{ccondprob}
\end{equation}

\begin{figure}[t]
\centering\includegraphics[width = .7 \linewidth,clip]{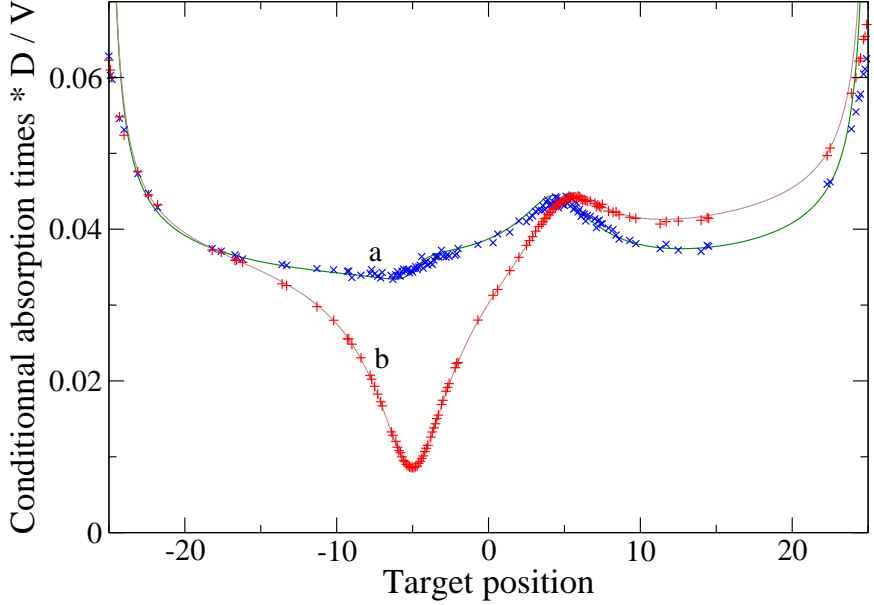}
\caption{(color online) 
Two-target simulations. The conditions are identical to those 
of Fig. \ref{c2t3D}; we show the conditional absorption times $\Tmm{1}$
(resp. $\Tmm{2}$). The blue Xs (resp. red +s) show the results of numerical
simulations; The green[a] (resp. brown[b]) solid line shows the theoretical 
estimation (\ref{ccondprob}) with the exact function $H$ for a sphere 
(\ref{exactsphere}).}
\label{ccondtime}
\end{figure}

We show in Fig.\ref{ccondtime} the result of numerical simulations. The noise
is more important than in other simulations, especially for $\Tmm{1}$. This is 
due to the fact that the probability $P_1$ is often small, which reduces the
number of processes on which the time is averaged, and thus increases the 
noise. 

We thus are able to compute first-passage times, splitting probabilities 
and absorption times with a good accuracy (especially with the improvements
given in Appendix \ref{refinements}), provided we know the pseudo-Green 
function $H$. The computation of $H$ is discussed extensively in Appendix 
\ref{approximations} and more briefly in the following. 

\section{Discussion \label{discussion}}

The computation of the pseudo-Green function can be a difficult problem. 
Indeed, there are a few cases when it can be computed exactly (see Appendix
\ref{exact}), namely in the discrete case for a rectangular/parallepipedic 
domain or for periodic boundary conditions, and in the continuous case 
when the domain is a disk, a sphere, or the surface of a sphere. 
Otherwise, we have to use an approximation, the simplest ones being presented 
and discussed in Appendix \ref{approximations}. 

In the following we present a synthetic and qualitative description of the
important parameters which have to be taken into account when it comes to 
computing the mean-first passage time.  

The first and most important parameter is the size of the domain. Indeed, 
the MFPT is proportional to the size of the domain, both in two and three 
dimensions. The second essential parameter is the size of the target for 
the continuous case: once we have these two parameters we already have a rough
order of magnitude of the MFPT. 
The third important parameter is the distance between the source and the 
target. In three dimensions this parameter is important as long as it is 
of the same order of magnitude as the target size; its influence is inversely 
proportional to the source-target distance. In two dimensions this parameter 
will be important at any distance, since the MFPT depends logarithmically of 
this distance. Once these parameter have been taken into account we have a 
good approximation of the MFPT (Eq. (\ref{base2D}) and (\ref{base3D})) if 
both source and target are far from any boundary. To see
what \emph{far} means in this case, a good criterion is that any correction
involving a boundary (see below) is negligible.  
Otherwise we have
an order of magnitude, and to proceed further we will have to take into 
account the precise position of the boundaries. 

The qualitative effect of the boundary is to increase the MFPT when the target
is near a boundary, and to decrease it when the source is near a boundary 
(it can be seen in the following equations). 
The first effect is much more important than the
second: in three dimensions, with a flat boundary, a basic approximation gives:
\begin{equation}
\Tm = \frac{1}{4\pi}\left(\frac1a - \frac1{|{\bf r}_S - {\bf r}_T|} 
+ \frac{1}{|{\bf r}_T - {\bf s}({\bf r}_T)|} - 
\frac1{|{\bf s}({\bf r}_S) - {\bf r}_T|}\right),
\end{equation}
where ${\bf s}({\bf r})$ denotes the point symmetrical to ${\bf r}$ with 
respect to the boundary. 
One can see that the influence of the boundary is inversely 
proportional to the distance between the target and the boundary. This is 
also true if the source is near a boundary, which is why the most important 
parameter is indeed the position of the target. One may note, however,  
that if the target or the source lies in a corner, these
effects are amplified.  

In two dimensions the influence of the position of the 
boundary is more important, and the position of the source is a relevant 
parameter: a basic approximation with a flat boundary gives: 
\begin{equation}
\Tm = \frac{1}{2\pi}\left(\ln\frac{|{\bf r}_S-{\bf r}_T|}{a} + 
\ln\frac{|{\bf s}({\bf r}_S)-{\bf r}_T|}{|{\bf s}({\bf r}_T)-{\bf r}_T|}\right)
\label{2Dboundary}
\end{equation}
If the target is much closer to the boundary than the source the effect 
can be to double the MFPT; on the other hand, if the source only is near 
a boundary, the related correction is bounded. The corners also have 
an amplifying effect in two dimensions. 

The quantitative estimates thus obtained are generally more accurate in 
three dimensions than in two dimensions, due to the fact that the effect of 
the boundaries on the pseudo-Green function is essentially local in three 
dimensions. In two dimensions there is still room for improvement, but an 
extensive discussion would be beyond the scope of this article.

\section*{Conclusion}

In this article we managed to compute the mean first-passage times, the 
splitting probability and the full probability density of the first-passage 
time
(in three dimensions) with a good accuracy for spherical or rectangular
domains. For other shapes (with a regular enough boundary), we gave the 
basic tools to approximately estimate these quantities. These results are 
especially important in the analysis of diffusion-limited reactions: The 
first-passage time corresponds to the reaction time if one of the reactants 
is static, and the reaction rate is infinite. 
Two promising extensions of our work would be to take into 
account \emph{finite} reaction rates, which would increase the relevance 
of our work to reaction-diffusion processes; and to study the same problem
with anomalous diffusion, which is relevant in many physical situations.

\section*{Acknowledgements}

We gratefully thank Jean-Marc Victor for useful discussions and comments, and
Sidney Redner for suggesting us the discrete simulation method.

\appendix

\section{\label{approximations}
Evaluation of the pseudo-Green function}

\subsection{Exact formulas\label{exact}}

\subsubsection{Periodic boundary condition and rectangular domains for a discrete 
pseudo-Green function}

There are two specific cases where the discrete pseudo-Green function $H$ 
may be computed exactly: when the domain is rectangular (parallepipedic in 
three dimensions) or when the boundary conditions are periodic \cite{CondaminJCP}. 
These results are interesting in themselves, but, moreover, for a domain which
is almost rectangular/parallepipedic, they will give a good approximation 
for $H$.
For periodic boundary conditions. if we consider a domain with $X$ sites in 
the $x$ direction, $Y$ sites in the $y$ direction, and $Z$ sites in the $z$ 
direction, a straightforward Fourier analysis gives: 
\begin{equation}
H({\bf r}|{\bf r}') = \frac1N \sum_{m=0}^{X-1}\sum_{n=0}^{Y-1}
\sum_{p=\delta_{(m,n)(0,0)}}^{Z-1}\frac{\exp\left(\frac{2im\pi(x-x')}{X}+
\frac{2in\pi(y-y')}{Y}+\frac{2ip\pi(z-z')}{Z}\right)}{1-\frac13\left(
\cos\frac{2m\pi}{X}+\cos\frac{2n\pi}{Y}+\cos\frac{2p\pi}{Z}\right)}
\end{equation}
In two dimensions, we have a similar formula for $H$: 
\begin{equation}
H({\bf r}|{\bf r}') = \frac1N \sum_{m=0}^{X-1}\sum_{m=\delta_{n0}}^{Y-1}
\frac{\exp\left(\frac{2im\pi(x-x')}{X}+
\frac{2in\pi(y-y')}{Y}\right)}{1-\frac12\left(
\cos\frac{2m\pi}{X}+\cos\frac{2n\pi}{Y}\right)}
\end{equation}

For a parallepipedic domain we get a slightly more complicated expression, 
and we have to use semi-integer coordinates for the points: 
$x$(resp. $y$ and $z$) varies between 
$1/2$ and $X$(resp. $Y$ and $Z$) $-1/2$. 
The result is the following: 
\begin{eqnarray}
H({\bf r}|{\bf r}') &=& \frac8N\sum_{m=1}^{X-1}\sum_{n=1}^{Y-1}\sum_{p=1}^{Z-1}
\frac{\cos\frac{m\pi x'}{X}\cos\frac{n\pi y'}{Y}
\cos\frac{p\pi z'}{Z}\cos\frac{m\pi x}{X}\cos\frac{n\pi y}{Y}
\cos\frac{p\pi z}{Z}}{1-\frac13\left(\cos\frac{m\pi}{X}+\cos\frac{n\pi}{Y}+
\cos\frac{p\pi}{Z}\right)} \label{exactcubic}\\
&& + \frac6N\sum_{m=1}^{X-1}\sum_{n=1}^{Y-1}
\frac{\cos\frac{m\pi x'}{X}\cos\frac{n\pi y'}{Y}
\cos\frac{m\pi x}{X}\cos\frac{n\pi y}{Y}
}{1-\frac12\left(\cos\frac{m\pi}{X}+\cos\frac{n\pi}{Y}
\right)} + 
\frac6N\sum_{p=1}^{Z-1}
\frac{\cos\frac{p\pi z'}{Z}
\cos\frac{p\pi z}{Z}
}{1-\cos\frac{p\pi}{Z}} \nonumber \\
&& + \frac6N\sum_{m=1}^{X-1}\sum_{p=1}^{Z-1}
\frac{\cos\frac{m\pi x'}{X}\cos\frac{p\pi z'}{Z}
\cos\frac{m\pi x}{X}\cos\frac{p\pi z}{Z}
}{1-\frac12\left(\cos\frac{m\pi}{X}+\cos\frac{p\pi}{Z}
\right)} + 
\frac6N\sum_{n=1}^{Y-1}
\frac{\cos\frac{n\pi y'}{Y}
\cos\frac{n\pi y}{Y}
}{1-\cos\frac{n\pi}{Y}} \nonumber\\
&& + \frac6N\sum_{n=1}^{Y-1}\sum_{p=1}^{Z-1}
\frac{\cos\frac{n\pi y'}{Y}\cos\frac{p\pi z'}{Z}
\cos\frac{n\pi y}{Y}\cos\frac{p\pi z}{Z}
}{1-\frac12\left(\cos\frac{n\pi}{Y}
+\cos\frac{p\pi}{Z}\right)} 
+ \frac6N\sum_{m=1}^{X-1}
\frac{\cos\frac{m\pi x'}{X}
\cos\frac{m\pi x}{X}}
{1-\cos\frac{p\pi}{Z}} \nonumber 
\end{eqnarray}
In two dimensions the expression is slightly less imposing:  
\begin{eqnarray}
H({\bf r}|{\bf r}')&=& \frac4N \sum_{m=1}^{X-1} \sum_{n=1}^{Y-1}
\frac{\cos\frac{m\pi x'}{X}\cos\frac{n\pi y'}{Y}
\cos\frac{m\pi x}{X}\cos\frac{n\pi y}{Y}}{
1-\frac12\left(\cos\frac{m\pi}{X}+\cos\frac{n\pi}{Y}\right)}\nonumber \\
&& + \frac4N \sum_{m=1}^{X-1}\frac{\cos\frac{m\pi x'}{X}\cos\frac{m\pi x}{X}}{
1 - \cos\frac{m\pi}{X}}  + 
\frac4N \sum_{n=1}^{Y-1}\frac{\cos\frac{n\pi y'}{Y}\cos\frac{n\pi y}{Y}}{
1 - \cos\frac{n\pi}{Y}} 
\end{eqnarray}

These formulae have the advantage of being exact, which enables us 
to compute exactly all the quantities studied in this article for such 
geometries. However, the computation of $H$ may be computationally expensive 
for large domains. 
In the continuous case, the same method can be applied, but $H$ can only 
be expressed as an infinite series\cite{Barton}. We give the result for a 
$2D$ rectangle $X \times Y$: 
\begin{eqnarray}
H({\bf r}|{\bf r}') & = & \frac4{XY} \sum_{m=1}^{\infty} \sum_{n=1}^{\infty}
\frac{\cos\frac{m\pi x'}{X}\cos\frac{n\pi y'}{Y}
\cos\frac{m\pi x}{X}\cos\frac{n\pi y}{Y}}{
\left(\frac{m \pi}{X}\right)^2+\left(\frac{n \pi}{Y}\right)^2}\nonumber \\
&& + \frac2{XY} \sum_{m=1}^{\infty}\frac{\cos\frac{m\pi x'}{X}\cos
\frac{m\pi x}{X}}{
\left(\frac{m\pi}{X}\right)^2}  + 
\frac2{XY} \sum_{n=1}^{\infty}\frac{\cos\frac{n\pi y'}{Y}\cos\frac{n\pi y}{Y}}{
\left(\frac{n\pi}{Y}\right)^2} 
\end{eqnarray}

\subsubsection{Disks and spheres for the continuous pseudo-Green functions \label{Disks}}

In the continuous case there is however a case where the pseudo-Green function
is known exactly: if the domain is a disk or sphere of radius $b$.
We will simply give the results; the detailed computation can be found 
in \cite{Barton}. 
\begin{figure}[t]
\centering\includegraphics[width = .5\linewidth,clip]{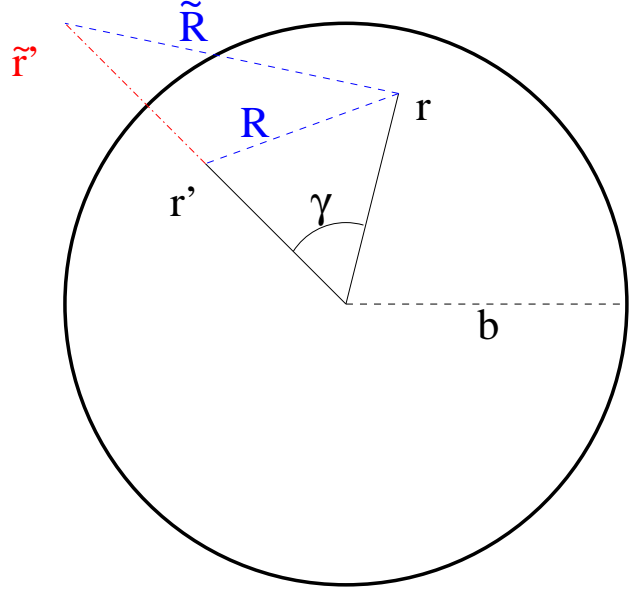}
\caption{(color online) 
Schematic picture of the quantities used in the computation of 
$H({\bf r}|{\bf r}')$.}
\label{configuration}
\end{figure}
In both formulas, we use the image of ${\bf r}'$, that we note 
$\tilde{\bf r}'$, which is aligned with ${\bf r}$ and the center of the 
disk/sphere $O$, and at a distance $\tilde{r}' = b^2/r'$. We note 
$R = |{\bf r}-{\bf r}'|$, $\tilde{R} = |{\bf r}-\tilde{\bf r}'|$, and 
$\mu = \cos\gamma$, $\gamma$ being the angle between ${\bf r}$ and ${\bf r}'$. 
In two dimensions, the result is the following: 
\begin{equation}
H({\bf r}|{\bf r}') = \frac1{2\pi}\left(\ln\frac{b}{R}+\ln\frac{b}{\tilde{R}}
+ \ln\frac{b}{r'}+\frac{r^2+r'^2}{2 b^2}\right)
\label{exactdisk}
\end{equation}
The first term corresponds to $G_0$, the second to the image of ${\bf r}'$, the
third term is needed to ensure the symmetry of $H$, and the last term 
corresponds to the $-1/V$ term in the definition of the pseudo-Green function. 
The three-dimensional result is a bit more complicated, with a logarithmic 
term whose physical signification is unclear: 
\begin{equation}
H({\bf r}|{\bf r}') = \frac{1}{4\pi}\left(\frac1R+\frac{b}{r'\tilde{R}} 
-\frac1b\ln\left(\frac{r'\tilde{R}}{b^2}+1-\frac{rr'\mu}{b^2}\right) 
+ \frac{r^2+r'^2}{2b^3}\right)
\label{exactsphere}
\end{equation}
These results are very useful by themselves, but they will also be useful to 
approximate $H$ near a curved boundary, as we will see in the following. The 
result for a sphere can also be used to estimate $\bar{H}$ when one uses the 
approximation $H=G_0$ in non-elongated 3D domains. Indeed the exact result 
enables one to take into account the corrections to $G_0$, which are negligible
when the source and the target are close, but give a substantial correction to 
the value of $\bar{H}$. To compute $\bar{H}$, one can use Eq. (\ref{4}), and 
choose for ${\bf r}$ the centre of the sphere. We have in this case: 
\begin{equation}
H({\bf r}'|{\bf r}=0) = \frac{1}{4\pi}\left(\frac1R+\frac{R^2}{2b^3}\right)
\end{equation}
A constant $(1 -\ln(2))/(4\pi b)$ 
has been suppressed, in order to have a final result
relevant for the approximation $H=G_0$. From this expression of $H$ it is 
straightforward to get an expression for $\bar{H}$: 
\begin{equation}
\bar{H} = \frac{3}{5}\left(\frac{3}{4\pi}\right)^{2/3}V^{-1/3}
\end{equation}
If one wants to use this result in the discrete case, it should be noted that 
the continuous limit of the discrete model corresponds to $D=1/2d$ and not 
$D = 1$. This diffusion coefficient is included in the discrete pseudo-Green 
function, and the discrete estimation of $\bar{H}$ is thus: 
\begin{equation}
\bar{H} = \frac{18}{5}\left(\frac{3}{4\pi}\right)^{2/3}N^{-1/3} 
\label{valuehbar}
\end{equation}

\subsubsection{Surface of spheres}

Another case where we can compute exactly $H$ is the case of the surface of 
a sphere. 
Indeed in this case we have exactly: 
\begin{equation}
H({\bf r}|{\bf r}') = -\frac1{2\pi}\ln|{\bf r}-{\bf r}'|
\end{equation}
Since $H$ is isotropic in this case it simplifies things: $G_0(a) + 
H^*({\bf r}_T|{\bf r}_T)$ can be replaced by $H(a)$ in Eq.(\ref{simplebrownian})
This gives back the result obtained by a straightforward computation of the FPTs in a sphere \cite{Desbois}. Moreover this
will give good approximations of all the two-target quantities, which was, to 
our knowledge, not known until now. This result is not used elsewhere 
in the paper, but is however important due to the physical relevance of the 
diffusion on the surface of a sphere.

\subsection{Use of the approximations\label{use}}

The next step is to study cases where no exact formula 
for $H$ is known. The simplest approximation to $H$ is the infinite-space 
Green function $G_0$, but this approximation in often unsatisfying. We thus 
present a few ways to improve it.  
Before we present them, it must be emphasised that, in general,
all the $H$ terms should be derived with the 
same approximation: $H$ is defined up to a constant, and this constant depends
of the approximation used! However, for complicated expression involving $H$, 
this constraint can be relaxed: if the expression can be decomposed into 
terms of the form $(H({\bf r}_1|{\bf r}_2) - H({\bf r}_3|{\bf r}_4))$, these
terms may be computed with different approximations, since they do not 
depend on the constant up to which $H$ is defined. 
For example, in the two-target problem, we have 
$P_1 = \frac{H_{1s}+H_{02}-H_{2s}-H_{12}}{H_{01}+H_{02}-2H_{12}} $. 
We can use if necessary two approximations, one accurate around $T_1$, 
which we note $H^{(1)}$, and another accurate around $T_2$, which we note 
$H^{(2)}$. Then, to compute $P_1$, we use them the following way: 
\begin{equation}
P_1 = \frac{H^{(1)}_{1s}+H^{(2)}_{02}-H^{(2)}_{2s}-H^{(1)}_{12}}{
H^{(1)}_{01}+H^{(2)}_{02}-H^{(1)}_{12}-H^{(2)}_{12}}
\end{equation}
This trick can be especially useful if one has to deal with two targets near 
two different boundaries. 

\subsection{Approximations\label{boundary}}

The most basic approximation is already known: it is the approximation 
$H = G_0$. Its physical meaning is to ignore the presence of the boundaries, 
as far as the pseudo-Green function is concerned. To improve this 
approximation, there are essentially two ways: the first is to take the 
boundaries into account locally, and to satisfy the boundary conditions at
the nearest boundary, we will see how in the following. The second one is to 
take the boundaries into account globally, by taking into account the terms 
$-1/N$ or $-1/V$ in the definition of $H$. The order of magnitude of the 
related correction will be of about $({\bf r}-{\bf r}')^2/N$ in the discrete 
case, or $({\bf r}-{\bf r}')^2/4A$ in the 2D continuous case, 
$({\bf r}-{\bf r}')^2/6V$ in the 3D continuous case. It is thus much 
weaker in 3D (the maximal relative correction scales as $N^{-1/3}$ or 
$a/V^{1/3}$) than in 2D (where the maximal relative correction scales as 
$1/\ln(N)$ or $1/\ln(V/a^3)$). A more detailed 
discussion of this kind of corrections would be technical and beyond the 
scope of this article, but the above order of magnitude can be a good 
evaluation of the accuracy of the following boundary approximation. 

This approximation takes explicitly into account a planar boundary, and 
ignores all the others. It can be used both in the continuous and in the 
discrete case.  
If we note ${\bf s}({\bf r})$ the point symmetrical to ${\bf r}$ with respect
to the boundary, then the local approximation: 
\begin{equation}
H({\bf r}|{\bf r}') = G_0({\bf r}-{\bf r}') + G_0({\bf r}-{\bf s}({\bf r}')) 
\label{Hlocal}
\end{equation}
satisfies the boundary conditions on the flat boundary, and is symmetric. 
It thus can be a good approximation for the pseudo-Green function. 

\begin{figure}[t]
\centering\includegraphics[width = .7\linewidth,clip]{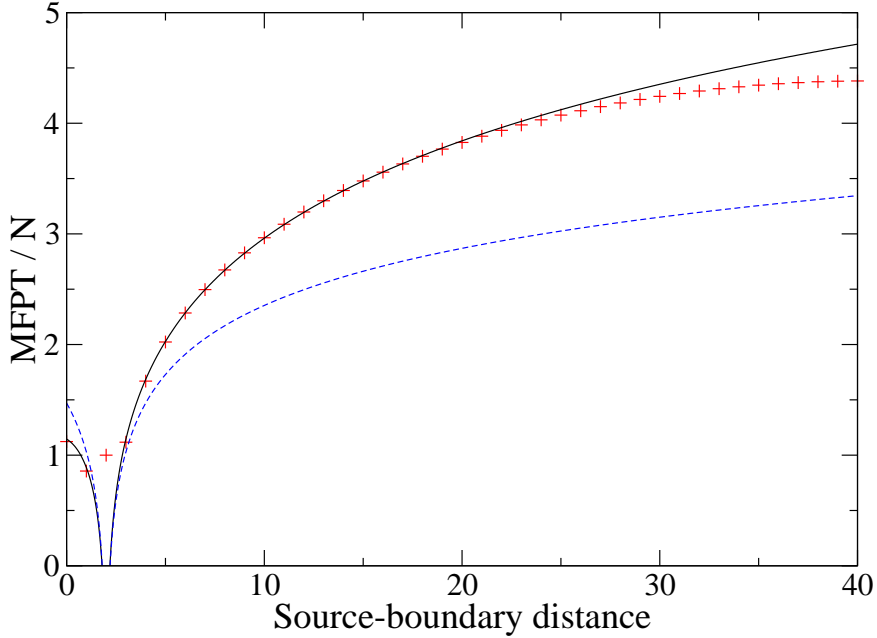}
\caption{(color online) 
Discrete random walk - Influence of the position of the source; 
the domain is a 2D square of side 41 centered on $(0,0,0)$, the
target is at $(-18,0)$ and the source is at $(x-20,0)$; blue dashed 
line: approximation $H=G_0$; black  solid line: local approximation taking into
account the boundary.}
\label{distsw2D}
\end{figure}

\begin{figure}[t]
\begin{minipage}{.7\linewidth}
\centering\includegraphics[width = \linewidth,clip]{quarter3D1}
\centering\includegraphics[width = .3\linewidth,clip]{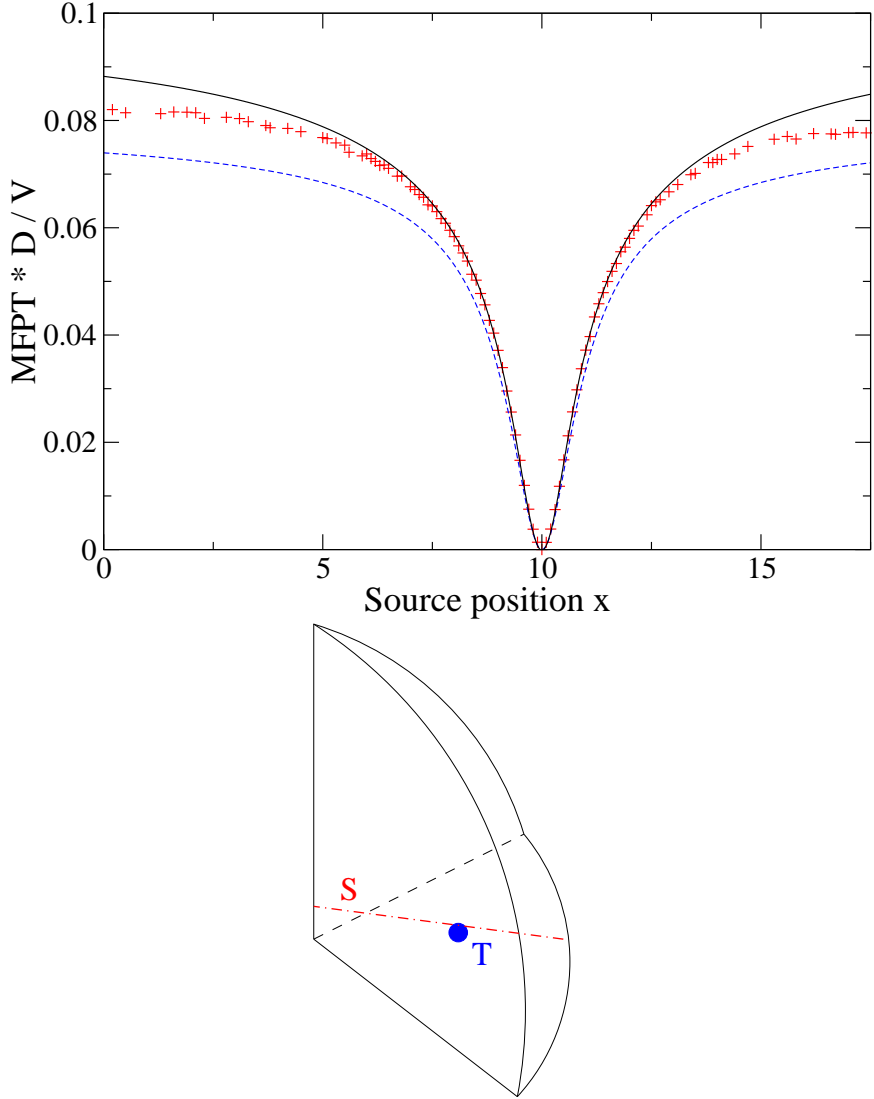}
\end{minipage}
\caption{(color online) 
3D Brownian motion: the domain is an eight of sphere; the sphere 
is of radius 25, centered on $(0,0,0)$, and the domain is reduced to positive
coordinates. The target is in $(10,10,2)$, and the source is in $(x,x,3)$.
Red crosses: numerical simulations;
blue dashed curve: basic approximation $H=G_0$; 
solid black curve: approximation (\ref{boundarybrownian}) with $H$ taking 
into account the nearest boundary (Eq. (\ref{Hlocal}))
}
\label{quarter3D1}
\end{figure}

Figs. \ref{distsw2D} and \ref{quarter3D1} show the efficiency of this 
approximation in two different cases: in a 2D discrete domain, and in a 3D 
continuous domain. In both cases the approximation improves the basic 
approximation $H=G_0$, but when the source is near another boundary, a 
systematic deviation appears, due to the influence of the other boundary. 

The curvature of the boundary may be taken into account by approximating 
the pseudo-Green function by the pseudo-Green function inside a circle 
(\ref{exactdisk}) or a sphere (\ref{exactsphere}), or outside a circle 
or a sphere (it can be found in \cite{Barton}).

\section{Computation of the higher-order moments\label{computation}}

\subsection{Discrete case\label{computationdiscret}}

In this part we will compute the higher-order moments of the FPT. 
 To do this, we start from an extension of Kac's formula  (see
Appendix \ref{AnnKac}), which is the relation between the Laplace
transforms of the FRT to the subset $\Sigma$, averaged on $\Sigma$, and
of the  FPT to this subset, the starting point being  averaged
over the complementary subset $\bar\Sigma$.

\begin{equation}
\pi(\Sigma)\left( \left\langle e^{-s\mathbf{T}}\right\rangle_\Sigma -
e^{-s} \right) =  (1 -\pi(\Sigma))\left(e^{-s} - 1\right) \left\langle
e^{-s\mathbf{T}} \right\rangle_{\bar{\Sigma}}
\label{HighKac}
\end{equation}

Both averages are weighted by the stationary probability $\pi$, in
the  following sense:
\begin{equation}
<\varphi(\mathbf{T})>_\Sigma = \frac1{\pi(\Sigma)} \sum_{i \in
\Sigma}\pi({\bf r}_i)  \sum_{t=1}^{\infty} p_i(\mathbf{T}=t) \varphi(t)
\end{equation}
\begin{equation}
<\varphi(\mathbf{T})>_{\bar\Sigma} = \frac1{1-\pi(\Sigma)} \sum_{i
\notin \Sigma}\pi({\bf r}_i)  \sum_{t=1}^{\infty} p_i(\mathbf{T}=t) \varphi(t),
\end{equation}
where $p_i(\mathbf{T}=t)$ is the probability for the FRT (or the FPT,
according to whether the point $i$ belongs to $\Sigma$ or not) to be
$t$, if the random walk starts from the point $i$. To apply the
equation (\ref{HighKac}) to the determination of the FPTs, we may
notice that the FPT from any point of the graph (except target) is
the same on the original graph and on the modified graph: indeed,  the
behaviour of a random walk is exactly the same on both lattices as
long as  they do not reach $T$, and what happens afterwards does not
matter. Moreover,  the FRT to $T$ is still the FPT from $S$ to $T$,
plus one.  Thus if we apply the formula (\ref{HighKac}) to the modified
graph, $\Sigma$ being reduced to $T$, we get the following relation between
the Laplace transform of the FPT from $S$ and the FPT averaged over the 
whole set of points (without $T$): 
\begin{equation}
J\left( \left<e^{-s\mathbf{T}}\right>_S - 1 \right) = 
(1-J)\left(1 - e^s\right) \left< e^{-s\mathbf{T}} \right>_{\bar\Sigma}
\end{equation}
$J$ is still $\pi({\bf r}_T)$. We have to pay attention to one thing: the 
average over $\bar\Sigma$ is weighted by the weights for the stationary 
distribution of the modified lattice. To go further we will have to consider 
\emph{all} the modified lattices with $T$ as target point, the starting 
point being any point of the set. 
We will denote $\pi_i$ the stationary distribution associated with the 
modified graph whose starting point is $i$, and $J_i=\pi_i({\bf r}_T) $. 
Thus, we may note: 
\begin{equation}
J_i\left( \left<e^{-s\mathbf{T}}\right>_i - 1 \right) = 
\left(1 - e^s\right) \sum_{j \neq T} \pi_i({\bf r}_j) 
\left< e^{-s\mathbf{T}} \right>_j
\end{equation}
From this, we may deduce the recurrence equation for the moments: 
\begin{equation}
\left<\mathbf{T}^n\right>_i = \frac1{J_i}\sum_{m=1}^n 
\sum_{j \neq T} (-1)^{m+1} {n \choose m} \pi_{i}({\bf r}_j)
\left<\mathbf{T}^{n-m}\right>_j
\end{equation}
We may thus compute explicitly the second moment. 
\begin{equation}
\left<\mathbf{T}^2\right>_i = \frac1{J_i}
\sum_{j \neq T} \pi_i({\bf r}_j)
\left( 2 \left<\mathbf{T}\right>_j - 1 \right)
\end{equation}

If we replace $\pi$ and $<\mathbf{T}>$ by their values, we get 
\begin{equation}
\left<\mathbf{T}^2\right>_i = \frac{2N}{J_i} \sum_{j \neq T}
\left(\frac{1-J_i}N + J_i H({\bf r}_j|{\bf r}_i) - J_i H({\bf r}_j|{\bf r}_T)
\right)\left(  H({\bf r}_T|{\bf r}_T)- H({\bf r}_T|{\bf r}_j)\right) 
- \frac{1-J_i}{J_i}
\end{equation}
We then may use the value of $\frac{1-J}J$, which we know: 
\begin{eqnarray}
\left<\mathbf{T}^2\right>_i & = & 2N  \sum_{j \neq T}
\left(H({\bf r}_T|{\bf r}_T)-H({\bf r}_T|{\bf r}_i) + H({\bf r}_j|{\bf r}_i) 
- H({\bf r}_j|{\bf r}_T) \label{exactt2}
\right)\left(  H({\bf r}_T|{\bf r}_T)- H({\bf r}_T|{\bf r}_j)\right)\nonumber
\\ && - N (H({\bf r}_T|{\bf r}_T)-H({\bf r}_T|{\bf r}_i))
\end{eqnarray}
This equation is exact, but it is difficult to evaluate
properly in the general case. We will thus use approximations to evaluate this 
expression in the case of a 3-D regular lattice, with $N$ large and the 
boundaries far from the target, at a typical distance $N^{\frac13}$. 
We can thus neglect the term 
$N (H({\bf r}_T|{\bf r}_T)-H({\bf r}_T|{\bf r}_i))$ in the R.H.S. of 
equation (\ref{exactt2}). If we develop the rest of the formula, we get: 
\begin{eqnarray}
\left<\mathbf{T}^2\right>_i & = & 2N \left(NH^2({\bf r}_T|{\bf r}_T) - N
H({\bf r}_T|{\bf r}_T)H({\bf r}_T|{\bf r}_i)+ H({\bf r}_T|{\bf r}_T)
\sum_{j \neq T}(H({\bf r}_j|{\bf r}_i)-2H({\bf r}_T|{\bf r}_j))\right.
\nonumber \\ && \left. + 
H({\bf r}_i|{\bf r}_T)\sum_{j \neq T}H({\bf r}_T|{\bf r}_j) 
- \sum_{j \neq T}H({\bf r}_T|{\bf r}_j)H({\bf r}_j|{\bf r}_i)
+ \sum_{j \neq T}H^2({\bf r}_T|{\bf r}_j) \right)
\end{eqnarray}
We can now drop the least important terms in this formula by evaluating the 
order of magnitude of the various sums over $j$.
We have (cf. Eq. (\ref{hbar})):
\begin{equation}
\frac1N \sum_{j} H({\bf r}_i|{\bf r}_j) = \bar{H}
\label{barh}
\end{equation}

Since $G_0({\bf r}) \sim 1/r$ in 3D, and the corrections are, on the worst 
case,of the same order of magnitude, we can see that $\bar{H}$ scales as 
$N^{-1/3}$. 
If we consider the sums $\sum_j H^2({\bf r}_T|{\bf r}_j)$ and 
$ \sum_j H({\bf r}_T|{\bf r}_j)H({\bf r}_j|{\bf r}_i)$, we may first notice 
that: 
\begin{equation}
\sum_j H({\bf r}_T|{\bf r}_j)H({\bf r}_j|{\bf r}_i) \leq 
\left(\sum_j H^2({\bf r}_T|{\bf r}_j)
\sum_j H^2({\bf r}_i|{\bf r}_j) \right)^{1/2}
\label{scaling}
\end{equation}
We thus only need to consider the case of 
$(1/N) \sum_j H^2({\bf r}_i|{\bf r}_j)$. 
And, for the same reasons as above, we can see that it scales as 
$N^{-2/3}$. 
Putting all this together, we have: 
\begin{equation}
\left<\mathbf{T}^2\right>_i = 2N^2\left[ \left(
H({\bf r}_T|{\bf r}_T)-H({\bf r}_T|{\bf r}_i)\right)
\left( H({\bf r}_T|{\bf r}_T) - \bar{H} \right)
+\mathcal{O}(N^{-2/3}) \right]
\end{equation}
It is possible to generalize this expression to higher-order moments; 
we will obtain the following result, for a given $n$: 
\begin{equation}
\left<\mathbf{T}^n\right>_i = n!N^n\left[ \left(
H({\bf r}_T|{\bf r}_T)-H({\bf r}_T|{\bf r}_i)\right)
\left( H({\bf r}_T|{\bf r}_T) - \bar{H} \right)^{n-1}
+\mathcal{O}(N^{-2/3}) \right]
\end{equation}
We can prove this by recurrence: 
if this is true for $m < n$, then: 
\begin{equation}
\left<\mathbf{T}^n\right>_i = \frac{n}{J_i}
\sum_{j \neq T}   \pi_i({\bf r}_j)
\left<\mathbf{T}^{n-1}\right>_j
\end{equation}
The others terms are negligible (their relative order of magnitude 
is at most $1/N$), and we will thus ignore them. 
We replace everything by its value, which gives:
\begin{equation}
\left<\mathbf{T}^n\right>_i = n!N^{n-1}\sum_{j \neq T}
\left(\begin{array}{l}
H({\bf r}_T|{\bf r}_T)-H({\bf r}_T|{\bf r}_i)\\
+H({\bf r}_j|{\bf r}_i)-H({\bf r}_j|{\bf r}_T)\\
\end{array}
\right)\left(\begin{array}{l}
H({\bf r}_T|{\bf r}_T)\\
-H({\bf r}_T|{\bf r}_j)\\
\end{array}\right)
\left(\begin{array}{l}
\left(H({\bf r}_T|{\bf r}_T)-\bar{H}\right)^{n-2}\\
+\mathcal{O}(N^{-\frac23}) \\
\end{array}
\right)
\end{equation}
Using exactly the same approximations as above (the computation is identical),
we get: 
\begin{equation}
\left<\mathbf{T}^n\right>_i = n!N^n\left[ \left(
H({\bf r}_T|{\bf r}_T)-H({\bf r}_T|{\bf r}_i)\right)
\left( H({\bf r}_T|{\bf r}_T) - \bar{H} \right)^{n-1}
+\mathcal{O}(N^{-2/3}) \right]
\end{equation}
As for the dependence with $n$ of the correction, since we perform exactly
the same operation at each step $n \rightarrow n+1$, the correction will be
proportional to $n$, which may help estimate the validity of the  
approximation. 

This computation fails for elongated domains: two main hypotheses are not 
satisfied in this case, namely that the boundaries are at a typical distance 
$N^{1/3}$, and that the corrections to $G_0$ have the same order of magnitude. 
The method can not either be applied to the 2D case, since the terms 
$1/N \sum_j H^2({\bf r}_i|{\bf r}_j)$ are no longer negligible. 

\subsection{Continuous case\label{computationcontinu}}

In the continuous case we can perform a similar computation. 
The higher-order moments of the FPT at the target
satisfy  the following equations\cite{Risken}:
\begin{equation} D \Delta \Tmur{n}{}
=-n\Tmur{n-1}{}\;{\rm if}\;
{\bf r}\in{\mathcal D}^*
\end{equation}
\begin{equation} \Tmur{n}{} = 0 \;{\rm if}\;
{\bf r} \in \Sigma_{\rm abs}
\end{equation}
\begin{equation}
  \partial_n\Tmur{n}{} = 0 \;{\rm if}\; {\bf r} \in \Sigma_{\rm refl}
\end{equation} Using a new time the Green function defined by Eqs.
(\ref{G1},\ref{G2},\ref{G3})
and the Green formula, we have
\begin{equation} \Tmur{n}{S} 
=\frac{n}{D}\int_{\mathcal{D}^*} G({\bf r}|{\bf r}_S)  
\Tmur{n-1}{} \rangle d^d{\bf r}
\end{equation}

With the knowledge of $G({\bf r}|{\bf r}')$ for all starting points 
${\bf r}$, it is possible to compute the full distribution. 
In three dimensions, it is possible to find an expression for 
$\left<\mathbf{T}^n\right>$ similar to the one found in the discrete case. 
We will start from Eq.(\ref{simplebrownian}). 
We can now compute the second moment, using the values for 
$\langle\mathbf{T}\rangle$ and $\rho_0$: 

\begin{eqnarray}
\Tmur{2}{S} & = & \frac{2V}{D^2}
\int_{\mathcal{D}^*}
\left[G_0(a)+ H^*({\bf r}_T|{\bf r}_T)-
H({\bf r}_T|{\bf r}_S) +H({\bf r}|{\bf r}_S) - H({\bf r}|{\bf r}_T)\right]\\
&& 
\left[G_0(a)+H^*({\bf  r}_T|{\bf r}_T)-H({\bf r}|{\bf r}_T)\right]
d^d{\bf r}
\nonumber
\end{eqnarray}

To compute this, we will use the two following equations, equivalent to Eq.
(\ref{barh}) and (\ref{scaling}) for discrete random walks: 
\begin{equation}
\int_{\mathcal{D}^*}H({\bf r}_0|{\bf r})d^d{\bf r} = V\bar{H}
+ \mathcal{O}\left(a^2\right)
\end{equation}
\begin{equation}
\int_{\mathcal{D}^*}H({\bf r}_1|{\bf r})H({\bf r}_2|{\bf r})
d^d{\bf r} = \mathcal{O}\left(V^{1/3}\right)
\end{equation}

This gives: 
\begin{equation}
\Tmur{2}{S} = \frac{2V^2}{D^2}\left[
\left(G_0(a) + H^*({\bf r}_T|{\bf r}_T) - H({\bf r}_T|{\bf r}_S)\right)
\left(G_0(a) + H^*({\bf r}_T|{\bf r}_T) - \bar{H}\right)
+\mathcal{O}\left(V^{-2/3}\right)\right]
\end{equation}
This result may be extended by recurrence to higher-order moments, in exactly
the same way that in the discrete case, which gives:
\begin{equation}
\Tmur{n}{S} = \frac{n!V^n}{D^n}\left[
\left(G_0(a) + H^*({\bf r}_T|{\bf r}_T) - H({\bf r}_T|{\bf r}_S)\right)
\left(G_0(a) + H^*({\bf r}_T|{\bf r}_T) - \bar{H}\right)^{n-1}
+\mathcal{O}\left(nV^{-2/3}a^{2-n} \right) \right]
\end{equation}

\section{\label{refinements}Refinements of the continuous theory}

In this Appendix we will see how to improve the results of Section \ref{continu}, 
provided we know the pseudo-Green function $H$. 
The results we obtained in Section  \ref{continu} are not perfectly satisfying 
for three reasons: 
\begin{itemize}
\item{When the source and the target are close, the approximation works 
better than one could naively expect, given that it does not satisfy
Eq. (\ref{G2}) very accurately. It would be interesting to understand why.}
\item{The approximation lacks accuracy when the target is near a boundary.}
\item{In the two-target case the accuracy is not very good when the two 
targets are close.}
\end{itemize}
We will treat the first point in detail, and give the corrections, and 
the method used to compute them, for the second and third point. 

\subsection{A better evaluation of $G$}

To understand this, we will first notice that the Green function we use could 
also be used in an electrostatic problem: the source is equivalent to a point 
charge, and the absorbing spheres are equivalent to conducting spheres set at 
a null potential. We can thus apply the well-known method of images 
\cite{Jackson} to our problem. 
If we have an image charge $q$  
\begin{equation}
q({\bf r}_S) = \left\{
\begin{array}{ll}
-\frac{a}{|{\bf r}_S-{\bf r}_T|}\equiv -\frac{a}{R}&\;{\rm in}\;{\rm 3D} \\
-1 & \;{\rm in}\;{\rm 2D} 
\end{array}\right.
\end{equation}
placed on ${\bf i}({\bf r}_S)$, located on the line between the center of the 
sphere and the source, at a distance $R'=a^2/R$ of the target, where $R$ is 
the source-target distance, then the solution: 
\begin{equation}
G({\bf r}|{\bf r}_S) = \rho_0({\bf r}_S) + G_0({\bf r}|{\bf r}_S) - 
G_0({\bf r}|{\bf r}_T) + 
q({\bf r}_S)(G_0({\bf r}|{\bf i}({\bf r}_S))-G_0({\bf r}|{\bf r}_T))
\end{equation}
satisfies exactly the boundary condition (\ref{G2}) on the target sphere: we 
have, for ${\bf r} \in \Sigma_{\rm abs}$ 
\begin{equation}
G_0({\bf r}|{\bf r}_S) - G_0({\bf r}|{\bf r}_T)
+ q({\bf r}_S)(G_0({\bf r}|{\bf i}({\bf r}_S))-G_0({\bf r}|{\bf r}_T)) = 
G_0({\bf r}_S|{\bf r}_T) - G_0(a)
\label{spherebc}
\end{equation}
However, this solution does not satisfy the reflecting boundary conditions, 
and we will rather use the solution: 
\begin{equation}
G({\bf r}|{\bf r}_S) = \rho_0({\bf r}_S) + H({\bf r}|{\bf r}_S) - H({\bf r}|{\bf r}_T)
+ \K{}{{\bf r}}{{\bf r}_S}
\end{equation}
which approximately satisfies (\ref{G2}), provided that we neglect the 
variations of $H^*({\bf r}|{\bf r}_S)$ and $H^*({\bf r}|{\bf r}_T)$ on the 
target sphere. 
With this approximation we get:  
\begin{equation}
\rho_0({\bf r}_S) = G_0(a) - H({\bf r}|{\bf r}_S) + H^*({\bf r}_T|{\bf r}_T)
+\Kx{}{{\bf r}_S}
\end{equation}
Note that the last term $\Kx{}{{\bf r}_S}$ can be neglected, since the 
variations of $H^*$ over the target sphere are neglected. Finally, to 
find the Eq. (\ref{simplebrownian}), the only condition is to neglect the
variations of $H^*$ over the target sphere, which will be a good 
approximation as soon as the target is far from any boundary. 
If this condition is satisfied, the approximation for the MFPT is accurate, 
even if the source is near the target. 

\subsection{Influence of a boundary}

\begin{figure}[t]
\centering\includegraphics[width = .5\linewidth,clip]{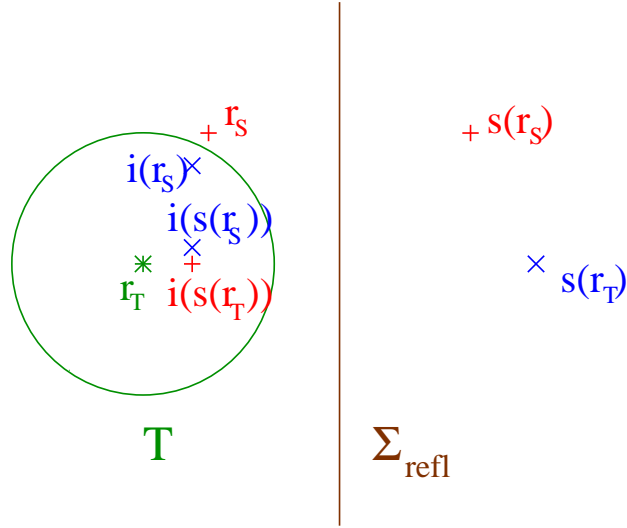}
\caption{(color online) 
Picture of the real and image charges when the target is near the 
boundary(+ = red plusses; - = blue crosses)}
\label{symetrie}
\end{figure}

If the target is near a boundary, however, $H^*$ can no longer be considered
as constant over the target sphere. 
To have a good approximation of $H$, one has to decompose the function one step
further: 
\begin{equation}
H({\bf r}|{\bf r}') = G_0({\bf r}|{\bf r}') + G_0({\bf r}|{\bf s}({\bf r}'))
+ H^{**}({\bf r}|{\bf r}'), 
\end{equation}
where ${\bf s}({\bf r})$ is the point symmetrical to ${\bf r}$ with respect to 
the boundary. This simply explicits the image charges due to the 
boundary, which themselves have images on the target sphere. 
The real and image charges are depicted in Fig. \ref{symetrie}. 

If we take into account all these charges, it is possible to obtain the 
following expression for the MFPT, valid as long as the target sphere does 
not touch the boundary:
: 

\begin{equation}
\Tmr{S} = \frac{V}{D}\left(
G_0(a) - H({\bf r}_T|{\bf r}_S) + H^*({\bf r}_T|{\bf r}_T) - K({\bf r}_S)
-K({\bf s}({\bf r}_S)) + K({\bf s}({\bf r}_T)) \right),
\label{boundarybrownian}
\end{equation}
where $K({\bf r}) = \Kx{}{{\bf r}}$.

\subsection{Two close targets}

The two-target case can be treated likewise: by considering the images of 
$T_1$ and $T_2$ on the other sphere, it is possible to compute corrections 
to the terms $H_{01}$, $H_{02}$, $H_{1s}$ and $H_{12}$ used in Eqs. 
(\ref{tcontinu}) and (\ref{pcontinu}). 
These correction are: 
\begin{equation}
H_{1s} = H({\bf r}_{T_1}|{\bf r}_S) + \K{2}{{\bf r}_{T_1}}{{\bf r}_S},
\label{c1}
\end{equation}
\begin{equation}
H_{01} = G_0(a_1) + H^*({\bf r}_{T_1}|{\bf r}_{T_1})
+ \K{2}{{\bf r}_{T_1}}{{\bf r}_{T_1}},
\label{c2}
\end{equation}
\begin{equation}
H_{12} = H({\bf r}_{T_1}|{\bf r}_{T_2}),
\label{c3}
\end{equation}
and similar corrections for $H_{02}$ and $H_{2S}$. $q_k({\bf r})$ and 
${\bf i}_k({\bf r})$ denote the value and the position of the image charge 
of ${\bf r}$ inside $T_k$.

\section{Proof of Kac's formula and of its extension}\label{AnnKac}

\subsection{The model}

We use the notations of Section \ref{discret}: 
$R$ is an arbitrary finite set of points $1,2, \hdots, N$, 
with positions ${\bf r}_1, {\bf r}_2, \hdots, {\bf r}_N$. 
$w_{ij}$ is the transition probability from $j$ to $i$, 
and we assume that any couple of points $i$ and $j$ in $R$ can be joined 
by at least one succession of links with non-zero transition probabilities. 

Among the points of $R$, we now arbitrarily define a subset $\Sigma$, and note
the complementary subset $\bar{\Sigma}$. 
Practically, the following properties 
will mostly be interesting if the number of points in $\Sigma$ is much smaller 
than the total number $N$ of points, but it is not necessary for the 
definitions. 

With the definitions, the Perron-Frobenius theorem \cite{VanKampen} 
assures that there exists a stationary probability $\pi({\bf r}_i)$,
which satisfies: 
\begin{equation}
\label{closed}
\pi({\bf r}_i) = \sum_{j\in R}w_{ij}\pi({\bf r}_j)
\end{equation}
From now on, we will consider that $\Sigma$ is absorbing, which means 
that the particle is absorbed as soon as it goes to the subset.  
However, it may start from it and go away on the following step 
without being absorbed. 
Thus, we state that, on any state ${\bf r}_i$, the particle has a 
probability $p_d({\bf r}_i)$ to be absorbed on its next step equal to:
\begin{equation}
p_d({\bf r}_i) = \sum_{j \in \Sigma}w_{ji}
\end{equation}

\subsection{Obtention of the formula}

Now, the probability $p({\bf r}_i,t)$ that
the conditional particle is adsorbed exactly at time $t$, starting from 
state $i$ at time $0$, obeys the backward equation: 

\begin{equation}
p({\bf r}_i,t) = \sum_{j \in \bar{\Sigma}} 
p({\bf r}_j,t-1)  
w_{ji}
\end{equation}
if $t \geq 2$, and
\begin{equation}
p({\bf r}_i,1) = p_d({\bf r}_i)
\end{equation}

As a result, the Laplace transform $\hat{p}$ of $p({\bf r}_i,t)$ satisfies: 

\begin{equation}
\hat{p}({\bf r}_i,s) -  e^{-s}\sum_{j \in \Sigma} 
w_{ji} = e^{-s} \sum_{j\in\bar{\Sigma}}
\hat{p}({\bf r}_j,s)  w_{ji},
\end{equation}

where $p({\bf r}_i,1)$ has been replaced by its value. 
We multiply this equation by the stationary probability $\pi({\bf r}_i)$
and sum up over all $i \in R$. We notice
that, from (\ref{closed})
\begin{equation}
\sum_{i\in\Sigma} w_{ji}\pi({\bf r}_i)  = \pi({\bf r}_j)
\end{equation}

We thus obtain: 

\begin{equation}
\label{prefin}
\sum_{i\in R} \hat{p}({\bf r}_i,s)
\pi({\bf r}_i) - e^{-s}\sum_{j\in\Sigma} \pi({\bf r}_j) = 
e^{-s} \sum_{j\in\bar{\Sigma}} 
\hat{p}({\bf r}_j,s) \pi({\bf r}_j)
\end{equation}

We now define two kinds of average for a quantity $\varphi(t)$:

({\it i}) the \emph{volume average}
\begin{equation}
\left< \varphi(\mathbf{T}) \right>_{\bar{\Sigma}} = 
\frac1{\pi(\bar{\Sigma})} \sum_{i\in\bar{\Sigma}} \pi({\bf r}_i)
\sum_{t=1}^{\infty} \varphi(t) p({\bf r}_i,t)
\end{equation}

({\it ii}) the \emph{surface average}
\begin{equation}
\left< \varphi(\mathbf{T}) \right>_{\Sigma} = 
\frac1{\pi(\Sigma)} \sum_{i\in\Sigma} \pi({\bf r}_i)
\sum_{t=1}^{\infty} \varphi(t) p({\bf r}_i,t)
\end{equation}

where $\pi(\bar{\Sigma})$ and $\pi(\Sigma)$ are the respective stationary 
probabilities of the volume and the surface: 
\begin{equation}
\pi(\bar{\Sigma}) = \sum_{i\in\bar{\Sigma}} \pi({\bf r}_i)
\end{equation}
\begin{equation}
\pi(\Sigma) = \sum_{i\in\Sigma} \pi({\bf r}_i), 
\end{equation}

and $\mathbf{T}$ denotes the absorption time, which corresponds to the 
FPT to $\Sigma$, or the FRT to $\Sigma$, depending on whether the starting 
point is on $\bar{\Sigma}$ or $\Sigma$. 
We thus simply get from (\ref{prefin}) the following equation: 
\begin{equation}
\pi(\Sigma)\lap{\Sigma} + \pi(\bar\Sigma)\lap{\bar{\Sigma}} - \pi(\Sigma)= 
e^{-s}\pi(\bar\Sigma)\lap{\bar{\Sigma}}
\end{equation}
or
\begin{equation}
\pi(\Sigma) \left(\lap{\Sigma} - e^{-s} \right) = \pi(\bar\Sigma ) \left( 
e^{-s} - 1 \right) \lap{\bar\Sigma },
\end{equation}
which is the extended Kac's formula, relating the Laplace transforms of 
the FRTs and the FPTs. 
Thus, for the first moment of $\mathbf{T}$ we obtain the very simple and 
general result: 
\begin{equation}
\left<\mathbf{T}\right>_{\Sigma} = \frac{1}{\pi(\Sigma)}
\end{equation}
(Kac's formula \cite{Aldous})

\section{Simulation methods}\label{simulations}

\subsection{Random walks}

For random walks we use a method based on the exact enumeration method
 \cite{Majid}. The exact enumeration method allows one to compute the 
exact distribution probability up to a given time: at each time step ($t >
0$), we  
compute the full probability distribution of the random walker, using the 
master equation: 
\begin{equation}
p({\bf r},t) = \frac1\sigma \sum_{{\bf r}' \in N({\bf r})} p({\bf r}',t-1)
\end{equation}
$p$ here is the probability of the random walker to be at position ${\bf r}$
at time $t$ and to never have reached the target site before. 
$N({\bf r})$ is the ensemble of neighbours of ${\bf r}$, which includes 
${\bf r}$ itself if ${\bf r}$ is a boundary site. The initial 
condition is of course $p({\bf r}, 0)=\delta({\bf r},{\bf r}_S)$. Note that 
if we set $T=S$ the algorithm will compute the distribution of the FRT.
After this first step, we have the probability distribution $p(t)$ of the FPT: 
\begin{equation}
p(t) = p({\bf r}_T, t)
\end{equation}
The last step of the algorithm
is to set $p({\bf r}_T,t)$ to 0, and we can then proceed to the 
computation for the time $t+1$. 
This enables us to compute the exact probability distribution, but of course 
the algorithm has to stop at a certain time. 
To go further, we can notice that the tail of the probability distribution 
is exponential (this corresponds to the highest eigenvalue of the transition 
matrix, the transition probabilities to and out of the target being set to 0
to take the absorption into account). 
If $p \sim e^{-\alpha t}$ for high enough $t$, then we can compute the
distribution up to a time $t_0$, then estimate:  
\begin{equation}
\Tm = \sum_{t=0}^{t_0-1}p(t)+\frac{p(t_0)t_0}{1-e^{-\alpha}}+
\frac{p(t_0)e^{-\alpha}}{\left(1-e^{-\alpha}\right)^2}
\end{equation}
The two latter terms correspond to $\sum_{t=t_0}^{\infty}p(t_0)e^{-\alpha(t-t_0)}$.
Since $\alpha$ is small, its order of magnitude being $1/N$, 
they are approximated by $p(t_0)t_0/\alpha + p(t_0)/\alpha^2$. 
To estimate $\alpha$, we take 
\begin{equation}
\alpha = \frac{1}{10}\ln\frac{p(t_0-10)}{p(t_0)}
\end{equation}
(we took $10$ steps and not one in order to avoid parity effects).
To select $t_0$, we run a few trial simulations, with a large maximum time, 
and we determine the minimal $t_0$ which gives a result differing by at most 
$0.1 \% $ from the result obtained with a larger $t_0$. We add a small 
security margin, and then run the simulation. 
We use similar methods for all the other quantities studied. 
The error on the simulation results is thus guaranteed to be less than 
$0.1 \% $! 

\subsection{Brownian motion}

Unfortunately, for the Brownian motion, we do not have such an accurate 
algorithm, and we thus used a Brownian-dynamics-based algorithm 
\cite{Bere97}: we average the time needed to reach the target on $n=10^5$ 
Brownian processes. To simulate the Brownian motion, we use the following 
algorithm: 
\begin{enumerate}
\item{Find the distance between the particle and the nearest obstacle (target,
non-flat boundary).}
\item{Multiply that distance by a constant $\alpha$ (we used $\alpha = 0.2$)
to get a trial typical step length.}
\item{If we are very close to a boundary, or very close to the target, 
this trial step length would be too small. We thus add a lower cutoff to this 
trial step length (we took $0.01$ near the target, of typical size radius 
$1$, and $0.2$ near the curved boundaries, whose radius of curvature was 
typically $25$), and get the typical step length $r_{step}$.}
\item{We use this step length to determine our time step 
$t_{step} = r_{step}^2$. 
(we have $D=1$).}
\item{For each direction $x,y,z$, we add to the position a Gaussian random 
variable, of variance $2t_{step}$. To get such a variable, we use two random 
variables $\nu$ and $\mu$
uniformly distributed between 0 and 1, and then the random variable $
r_{step} \sqrt{-2\ln(\nu)}\cos(2\pi\mu)$ is indeed a Gaussian with the required
variance.}
\item{If we are outside the domain, we move the particle inside the domain, 
to a position symmetrical with respect to the boundary.}
\item{If we are inside the target, we end the process, otherwise we 
take another step.}
\end{enumerate} 
This algorithm is less accurate than the one we used in the discrete case, and 
is computationally more expensive. Moreover, the study of the probability 
density of the FPT is delicate with this algorithm. 

\section{Properties of the pseudo-Green function $H$}\label{AnnGreen}

The properties of the \emph{continuous} Pseudo-Green function are well 
described in \cite{Barton}, and we will just describe the properties of 
the discrete one. We consider the case of symmetric transition 
probabilities. 
We define the discrete Laplacian operator: 
\begin{equation}
(-\Delta)_{ij} = \delta_{ij} - w_{ij} 
\end{equation}
This operator is hermitian, which will be useful. 
We define $\Phi_p$ and $\lambda_p$ the eigenvectors and (real) eigenvalues of 
the 
operator $-\Delta$, ordered from $0$ to $N-1$ in increasing order. We have
$\lambda_0 = 0$, and $\Phi_0 = 1/\sqrt{N}$, with the usual normalization. 
Since the operator is hermitian, 
we can take $\Phi_p^* = \Phi_p$. 
 We define: 
\begin{equation}
H({\bf r}_i|{\bf r}_j) = \sum_{p = 1}^{N-1} 
\Phi_p^*({\bf r}_j)\Phi_p({\bf r}_i)/\lambda_p
\end{equation}
This solution satisfies: 
\begin{equation}
-\Delta H({\bf r}_i|{\bf r}_j) = \delta_{ij} - \frac1N,
\end{equation}
which corresponds to the definition we used for $H$, and we thus found the 
solution (up to a constant) to the equation (\ref{pseudogreen}) we 
used to define $H$. This shows that $H$ is symmetric in its arguments if 
$W=\{ w_{ij} \}$ is symmetric. 
To prove that the sum $\bar{H}_j = \frac1N \sum_{i=1}^N 
H({\bf r}_i|{\bf r}_j)$ is independent of $j$, we will simply sum up the 
equation (\ref{pseudogreen}) over all $i$, and use the fact that $H$ is 
symmetric.
This gives: 
\begin{equation}
-\Delta \bar{H}_j = 0
\end{equation}
and $\bar{H}$ is proportional to $\Phi_0$, and thus is a constant.

\end{document}